\documentclass[manuscript]{acmart}

\usepackage{tabularx}
\usepackage{multirow}
\usepackage{soul}
\usepackage{array, makecell}
\usepackage{colortbl}
\usepackage{graphicx}

\AtBeginDocument{%
  \providecommand\BibTeX{{%
    \normalfont B\kern-0.5em{\scshape i\kern-0.25em b}\kern-0.8em\TeX}}}

\setcopyright{acmcopyright}
\copyrightyear{2024}
\acmYear{2024}
\acmDOI{XXXXXXX.XXXXXXX}

\acmConference[ACM FAccT '24]{Make sure to enter the correct
  conference title from your rights confirmation emai}{June 03--06,
  2024}{Rio de Janeiro, Brazil}
\acmBooktitle{ACM FAccT '24: ACM Conference on Fairness, Accountability, and Transparency,
 June 03--06, 2024, Rio de Janeiro, Brazil} 
\acmPrice{15.00}
\acmISBN{978-1-4503-XXXX-X/18/06}

\begin{document}
\newcommand{\ant}{anthropomorphization}

\title[From ``AI'' to Probabilistic Automation PREPRINT]{From ``AI'' to Probabilistic Automation: How Does Anthropomorphization of Technical Systems Descriptions Influence Trust?}





\author{Nanna Inie}
\orcid{0000-0002-5375-9542}
\affiliation{%
  \institution{IT University of Copenhagen, Center for Computing Education Research (CCER)}
  \streetaddress{Rued Langgaards Vej 7}
  \city{Copenhagen}
  \country{Denmark}}
\affiliation{%
 \institution{University of Washington, Paul G. Allen School of Computer Science \& Engineering}
 \streetaddress{3800 E Stevens Way NE}
 \city{Seattle}
 \country{USA}}
\email{nans@itu.dk}

\author{Stefania Druga}
\orcid{}
\affiliation{%
  \institution{University of Chicago, Center for Applied AI Research}
  \streetaddress{5807 S. Woodlawn Ave.}
  \city{Chicago}
  \country{USA}}
\email{stefania@hackidemia.com}

\author{Peter Zukerman}
\affiliation{%
  \institution{University of Washington, Department of Linguistics}
  \streetaddress{Guggenheim Hall}
  \city{Seattle}
  \country{USA}}
\email{pzukerman17@gmail.com}

\author{Emily M. Bender}
\affiliation{%
  \institution{University of Washington, Department of Linguistics}
  \streetaddress{Guggenheim Hall}
  \city{Seattle}
  \country{USA}}
\email{ebender@uw.edu}

\renewcommand{\shortauthors}{Inie et al.}

\begin{abstract}
  This paper investigates the influence of anthropomorphized descriptions of so-called ``AI'' (artificial intelligence) systems on people's self-assessment of trust in the system. Building on prior work, we define four categories of anthropomorphization (\textit{1. Properties of a cognizer}, \textit{2. Agency}, \textit{3. Biological metaphors}, and \textit{4. Properties of a communicator}). We use a survey-based approach ($n$=954) to investigate whether participants are likely to trust one of two (fictitious) ``AI'' systems by randomly assigning people to see either an anthropomorphized or a de-anthropomorphized description of the systems. We find that participants are no more likely to trust anthropomorphized over de-anthropmorphized product descriptions overall. The type of product or system in combination with different anthropomorphic categories appears to exert greater influence on trust than anthropomorphizing language alone, and \textit{age} is the only demographic factor that significantly correlates with people's preference for anthropomorphized or de-anthropomorphized descriptions. When elaborating on their choices, participants highlight factors such as \textit{lesser of two evils}, \textit{lower or higher stakes contexts}, and \textit{human favoritism} as driving motivations when choosing between product A and B, irrespective of whether they saw an anthropomorphized or a de-anthropomorphized description of the product. Our results suggest that ``anthropomorphism'' in ``AI'' descriptions is an aggregate concept that may influence different groups differently, and provide nuance to the discussion of whether anthropomorphization leads to higher trust and over-reliance by the general public in systems sold as ``AI''.
\end{abstract}




\received{20 February 2007}
\received[revised]{12 March 2009}
\received[accepted]{5 June 2009}

\maketitle

\section{Introduction}
Anthropomorphism, or the attribution of human characteristics or behavior to inanimate objects, is a common sense-making practice for people. With the advent of more advanced technical systems, anthropomorphism is often used to describe technical products (i.e. ``A.I. Shows Signs of Human Reasoning'' \cite{MicrosofNYT:ai_reasoning}), and it appears a rising trend in news coverage \cite{bunz2022ai}. This phenomenon --- anthropomorphizing\footnote{In this paper, we use the term \textit{anthropomorphization} when describing the intentional act of `putting anthropomorphic features into a product' or `using anthropomorphic words to describe a product'. The creator of the product or the writer of the text is responsible for the anthropomorphization, whereas \textit{anthropomorphism} denotes the process internal for the perceiver or user when human qualities are attributed to the system \cite{seeger2021texting}.} technical systems --- has been criticized for setting the wrong expectations and causing over-reliance in technology \cite{shanahan2022talking, shardlow2022deanthropomorphising, hunger2023unhype}.

Emily Tucker, the Executive Director at the Center on Privacy \& Technology at Georgetown Law, wrote in her 2022 Medium post \textit{Artifice and Intelligence} a declaration of intention to stop using the words ``Artificial intelligence'', ``AI'', and ``machine learning'' for the purpose of exposing and mitigating harms of digital technologies to individuals and communities, based on the underlying risk that the public will assume that ``AI'' technologies are more capable than they are \cite{tucker2022artifice}. Francis Hunger \cite{hunger2023unhype} also argues that \textit{``the use of anthropomorphising language is fueling AI hype. [It]] is problematic since it covers up the negative consequences of AI use.''} The argument here is that by using personified language when referring to AI systems, we also implicitly attribute human-like properties to them, which both makes them seem more powerful than they are while obscuring their potential negative effects. 

Prior studies investigated how conceptual metaphors influenced people's perception of algorithmic decision-making systems more broadly \cite{langer2022look}, as well as how anthropomorphic cues influence people's trust in robots \cite{christoforakos2021can}, voice-assistants \cite{fetterolf2022s}, and websites \cite{sah2015effects}. To our knowledge, no study has yet investigated anthropomorphic descriptions of products and systems powered by ``AI'', which we will refer to as \textit{probabilistic automation} systems,\footnote{The denomination ``artificial intelligence'' is poorly defined, and does not refer to a coherent set of technologies. In general, we find that discussions of technologies called ``AI'' become more lucid and thus productive when we speak about the automation of specific tasks. In the case of this research, the fictitious systems presented to our participants vary in their task domain, but they  are all imagined to be built on statistical analysis of large datasets. Therefore, we will refer to these systems collectively as ``probabilistic automation''.} influence people's trust and desire to use such systems. This motivated us to explore the overall research question: ``What are the effects on trust of anthropomorphization of probabilistic automation systems?''

We are specifically concerned with what we will call ``anthropomorphization by description'', rather than anthropomorphization by design\,---\,meaning we investigate the language used to \textit{describe} systems, rather than the language (and other attributes) built into the systems themselves. Whereas both types could have negative consequences, anthropomorphization by description is especially relevant in public discourse, where journalists, politicians, and copy-editors carry a significant responsibility for the use and spread of metaphors and analogies that will shape the public perception. 



We use a survey-based approach ($n$=954) to investigate whether participants believe themselves to be more likely to trust one of two (fictitious) probabilistic automation systems. Our investigation makes three contributions: First, we provide empirical evidence that people are \textit{not} more likely to choose anthropomorphized descriptions of products over de-anthropomorphized descriptions. Second, we find that some product types in \textit{combination} with different categories of anthropomorphizing language appear to have more influence on trust than anthropomorphizing language alone. Finally, we find that \textit{age} is the only variable that seems to have a dependent association with preferences
for anthropomorphized/de-anthropomorphized product descriptions.

\section{Background and related work}

\subsection{Metaphors, anthropomorphism, and technology}

Language shapes our interactions with technology. Even short textual descriptions can influence how humans meet and evaluate digital systems \cite{hartmann2008framing, strait2018robots, khadpe2020conceptual, langer2022look, kim2023communicating, kim2020effect}. In the context of probabilistic automation systems, the conceptual metaphor \cite{mcglone1996conceptual, crawford2009conceptual} or ``pitch'' of the system's functionality might play an especially compelling role, given the complexity of such systems \cite{langer2022look}. Accurately priming the user and adjusting their expectations to the system is difficult, and simply providing performance metrics is not meaningful to the average user, given their lack of familiarity with the inner workings of the technologies that they interact with~\cite{kocielnik2019will, khadpe2020conceptual}. In the absence of technical understanding, humans develop their own simplified mental models of how a system works --- models that are not always consistent with the actual functionalities of the system, and of which inaccurate versions can lead to consequences from mundanely inconvenient to more severe \cite{norman2013design}. 

Research on human interactions with technological devices shows a clear tendency of anthropomorphism. For example, humans are capable of engaging socially with machines \cite{isbister2000consistency,nass2000machines,reeves1996people}. This is especially true of robots and embodied assistants \cite{fong_survey_socially_interactive_bots,kanda2004interactive, tanaka2007socialization}. The more life-like a probabilistic automation application is in terms of embodiment (the physical form of the system), physical presence, social presence, and appearance, the more persuasive it can become \cite{bainbridge2008effect,roubroeks2011artificial}. For example, Vollmer et al. showed that robots could even exert peer pressure over children \cite{vollmer2018children}. In their experiment, 7- to 9-year-old children had a tendency to echo the incorrect, but unanimous, responses of a group of robots to a simple visual task \cite{vollmer2018children}. Smart voice assistants also lead children to overestimate the intelligence of these devices, trusting them, and deferring to them when making decisions \cite{druga2017hey}.

\subsection{Risks associated with anthropomorphization}

With the blight of publicly-available Large Language Models (LLMs) and generative probabilistic automation technology, numerous academic papers have appeared which warn about the risks of overusing anthropomorphic language to describe such technology \cite{salles2020anthropomorphism, shardlow2022deanthropomorphising, shanahan2022talking, abercrombie2023mirages, hunger2023unhype, deroy2023ethics}. Previous research has raised several categories of (interrelated) risks of anthropomorphization, detailed briefly below.

\subsubsection{Misplaced trust and over-reliance}
One direct consequence of anthropomorphization is misplaced trust, which in turn can lead to over-reliance on probabilistic automation systems \cite{salles2020anthropomorphism, hunger2023unhype, desai2023metaphors, abercrombie2023mirages, deroy2023ethics}. While anthropomorphism may enhance user experience and trust (in fact, much of the literature on anthropomorphism and technology concerns using anthropomorphization to increase trust, e.g., \cite{chen2021anthropomorphism, kim2021should, chi2023investigating}),\footnote{It should go without saying, but we note for good measure nonetheless that to seek to increase trust rather than trust-worthiness is to court risk.} it also risks creating a false sense of the system's capabilities. 
Such misplaced trust can be particularly problematic in high-stakes scenarios, such as medical diagnosis or financial decision-making, where over-reliance on probabilistic automation can lead to significant consequences. 

\subsubsection{Spillover effect of cognitive overestimation}
When probabilistic automation is perceived as having advanced cognitive properties, users may overestimate its capabilities in areas not directly demonstrated \cite{desai2023metaphors, abercrombie2023mirages}. For instance, if an probabilistic automation system is adept at data processing and pattern recognition, users might erroneously assume it is equally proficient in complex decision-making or ethical judgments. This cognitive overestimation can result in the inappropriate application of probabilistic automation advice, potentially leading to harmful outcomes. 


\subsubsection{Transparency and accountability}
When probabilistic automation systems are perceived as autonomous agents, it raises complex questions about accountability \cite{hunger2023unhype, bigman2019holding, stuart2021guilty}. In cases of error or malfunction, determining responsibility can be challenging, especially when users have been led to view these systems as `intelligent' entities. Some research has shown that people are aware of the dangers of overattributing accountability to
technology `when harm comes to pass' \cite{stuart2021guilty}, but the dynamics are not well understood.


Though there are many good arguments for \textit{not} anthropomorphizing probabilistic automation systems and not many good arguments for doing so, there are few scientific explorations of the details of anthropomorphic language and its specific impact. Our goal with this research was to take a first step towards understanding the phenomenon of anthropomorphization better.

\section{Methodology}

We designed our experiment to address the following \textbf{research questions}:
\begin{enumerate}
    \item Are people more likely to trust products that are described in anthropomorphizing language than products which are \textit{not} described in anthropomorphizing language?
    \begin{enumerate}
        \item Are people more likely to trust anthropomorphized products if imagining themselves as a user (\textit{personal trust}) than to trust them in use for the general population (\textit{general trust})?
    \end{enumerate}
    \item Are people more likely to trust products when the products are described in different \textit{kinds} of anthropomorphizing language? 
    \item Are different groups of people more likely to trust products that are described in anthropomorphizing language? (We investigated the groups gender, age, socio-economic status, level of education, and level of computer knowledge\footnote{Abercrombie et al. suggested that negative impacts of anthropomorphization could be exacerbated in ``vulnerable populations'' \cite{abercrombie2023mirages}.}). 
\end{enumerate}

\subsection{Defining anthropomorphic language} \label{sec:antcategories}

To investigate the influence of anthropomorphic language, we need to create a working definition of what that language is. In general, anthropomorphization is the assigning of human characteristics to non-human entities. Examining previous literature, we identified four general classes of anthropomorphizing language:

\begin{enumerate}
    \item Using predicates that portray the machine as a \textit{\textbf{cognizer}} \cite{desai2023metaphors, deroy2023ethics, abercrombie2023mirages, tucker2022artifice, salles2020anthropomorphism, isaeva2022anthropomorphism}. The human characteristic that seems most salient in the context of probabilistic automation is cognition: the ability to perceive, think, reflect, and experience things --- often expressed with the word `intelligent' or `intelligence'. Algorithms being anthropomorphized with \textit{\textbf{Properties of a cognizer}} might \textit{know, believe} or \textit{decide}.
    \item Describing the machine as an \textit{\textbf{agent}} \cite{hunger2023unhype, isaeva2022anthropomorphism} of an action. Hunger \cite{hunger2023unhype} posed anthropomorphization of a category she called `Active verbs', but we specify this slightly to include some degree of intention or independence, since machines can actively process many things without being attributed human capabilities. We therefore called this category \textit{\textbf{Agency}}. Those being anthorpomorphized in this category \textit{collect, monitor}, or \textit{choose}.
    \item Another category is using \textit{\textbf{Biological metaphors}} \cite{hunger2023unhype, salles2020anthropomorphism} to describe computational concepts. Those being anthropomorphized through biological metaphors might comprise \textit{neural nets} or have \textit{neurons} and \textit{synapses}.
    \item Finally, using verbs of \textit{\textbf{communication}} \cite{shardlow2022deanthropomorphising, isaeva2022anthropomorphism}. Those being anthropomorphized via \textit{\textbf{Properties of a communicator}} might be \textit{asked} things by users and \textit{tell} the user things in return.
\end{enumerate}   

\noindent These boundaries overlap somewhat: A computer being described as \textit{deciding} is both being cast in an agentive role and as a cognizer. Similarly, if a machine is said to \textit{see} something, that is both a biological metaphor and an attribution of cognition, and so on. We also don't expect these categories to fully cover all the ways that we use language to anthropomorphize algorithms. To get a sense of whether they cover a significant amount, however, we selected a text to annotate for anthropomorphizing language. Three of the authors independently annotated these texts, and used them as source of discussion before writing our own product descriptions (which all authors contributed to).

As one means of defining whether language is anthropomorphizing or not, we accessed the FrameNet database \cite{baker1998berkeley}. This resource describes words in terms of the \textit{frames} they describe and the \textit{frame elements} that participate in the frames. For example, the word \textit{imagine} expresses the Awareness frame, with frame elements Cognizer, Content, Topic and Element. We used the notion of the Cognizer frame element to look up words in the FrameNet resource which portray one of their arguments as a Cognizer. If the computational system is filling this role, then it is being anthropomorphized by having cognition attributed to it. Similarly, to assess words used of the \textit{Communication}-category, we looked up words related to the frame Communicator.

\subsection{Participants and recruitment}
Participants were recruited via the data collection platform Prolific, and compensated between £9-£15/h\footnote{As suggested by Prolific's standards for ``Good hourly rate''.} (depending on their average time to completion)  for their participation. This database allowed us to create pre-screening criteria such as country of residence, self-assessed socio-economic status, and ethnicity, to reach as diverse a group as possible (see Section \ref{sec:demographics} in the appendix for demographics). All participants signed a consent form that their (anonymous) answers could be used for research purposes.

\subsection{Experiment design}
We imagined eight pairs of fictional products based on some form of (relatively vague) probabilistic automation technology, giving 16 products total. For each product, we wrote a short ``pitch'' (less than 80 words), briefly describing the features of the product (the descriptions can be found in the appendix, Tables \ref{tab:productdescriptions1}- \ref{tab:productdescriptions4}). The goal of these pitches was to give a sense of the functionality of the product without being more technical than one would expect in a news article or popular literature description of a product. The products were paired in genres, so they would be somewhat comparable (for instance, ``recommender systems'' or ``online health diagnostics''), to enable apples-to-apples comparisons. The participant would always be asked to choose between product A and product B in one of the genres, and never between, e.g., an autonomous vehicle and a tutoring app. An overview of the products is shown in Table \ref{tab:products}. For each product, we wrote an anthropomorphized short pitch, and a de-anthropomorphized short pitch. The participants were randomly shown a combination of either:

\begin{quote}
    [Product A: Anthropomorphized description] + [Product B: De-anthropomorphized description] \textbf{or}
    [Product A: De-anthropomorphized description] + [Product B: Anthropomorphized description]
\end{quote}

\noindent and asked to choose between the two with one of the following questions:

\begin{itemize}
    \item \textbf{Thinking of yourself as a user, which of these systems are you more likely to trust?} We ask you to think about how likely you would be to trust using this system for your own purposes, assuming you would like to use the service it would provide (\textit{personal trust}).
    \item \textbf{Which of these systems do you think would give better output for its users?} Where ``better output'' means, for instance, more correct or more helpful output (\textit{general trust/reliance}).
\end{itemize}

\noindent ``Trust'' is inherently difficult to evaluate independent of context, but giving participants two options to choose between (`joint evaluation') has been shown to make it easier for people to evaluate ``difficult-to-evaluate attributes'' \cite{hsee1999preference, hsee1998will}. The questions were designed to reflect two essential questions for measuring trust identified by Hoffman et al. \cite{hoffman2018metrics} with two modifications: (1) We could not ask the user to evaluate the system's output (question 2 in \cite{hoffman2018metrics} addresses \textit{reliance} of output), given that the system does not exist in reality. We therefore created a distinction between \textit{personal trust} and \textit{general trust}. (2) To make it more likely that participants would understand trust in a somewhat similar way, an introductory text as well as a short definition of trust was provided with each product pair (see appendix, Section \ref{sec:surveyscreenshot}).

For each presentation of a product pair to a participant, we randomized which product of the pair would be presented in its anthropomorphize guise, and which de-anthropomorphized, but there was always one of each,  
and all participants were presented with all eight product pairs. Under each choice of product pair, we included an optional open answer text field where the participant could elaborate on their answer if they wanted to. A screenshot of the survey as it was presented to a participant is included in Section \ref{sec:surveyscreenshot}, Figure \ref{fig:surveyscreenshot} in the appendix.

\begin{table} 
    \centering
    \renewcommand{\arraystretch}{1.1}%
    \small
    \begin{tabular}{l|m{2cm}|m{2cm}|m{2.7cm}|m{2.7cm}}
        \textbf{Product genre} & \textbf{Product A} & \textbf{Product B} & \textbf{Category, Study 1} & \textbf{Category, Study 2} \\ \hline
        \textbf{Recommender systems} & re-Commender & IntelliTrade & Cognizer & Agency \\ \hline
        \textbf{Personal assistant} & MonAI Maker & Cameron & Cognizer & Agency \\ \hline
        \textbf{Autonomous vehicles} & HaulIT & Commuter & Agency & Biological metaphors \\ \hline
        \textbf{Drones} & AquaSentinel & AI Scan Guards & Agency & Biological metaphors \\ \hline
        \textbf{Legal recommendations} & Judy & JurisDecide & Biological metaphors & Communicator \\ \hline
        \textbf{Online health diagnoses} & MindHealth & DermAI Scan & Biological metaphors & Communicator \\ \hline
        \textbf{AI Tutor} & Lingua & MentorMe & Communicator & Cognizer \\ \hline
        \textbf{Assisted shopping} & WardrobEase & Shoppr & Communicator & Cognizer \\
    \end{tabular}
    \vspace{2mm}
    \caption{Overview of the different probabilistic automation-based products and their genres.}
    \label{tab:products}
\end{table}

\subsection{Survey design}

The survey was created in the software SurveyXact. For the initial development of the pitches (as used in the Pilot and Study 1), we arbitrarily assigned the product pairs to one of the anthropomorphic language categories defined in section \ref{sec:antcategories}. In Study 2, we arbitrarily ``swapped'' anthropomorphization categories between the product pairs, to avoid overinterpretation of results based on one study alone\,---\,see Table \ref{tab:products}.

\subsubsection{Pilot study}
We ran a pilot study with 37 participants recruited through personal networks. Only minor edits to the product descriptions were made to clarify misunderstandings as a result of the pilot study. 

\subsubsection{Study 1}
For Study 1, 333 participants signed the consent form, and 313 participants completed the survey fully,  while 20 participants partially completed the survey. We have included all partially completed survey responses in the analyses, as they provide valid answers to the questions. Excluding these participants has no statistically significant impact on the results. Participants were asked about both \textit{personal trust} and \textit{general trust}, meaning that for each product pair, they were asked to evaluate which product they would be more likely to trust for themselves as a user, and subsequently (but visible on the same page), which product they believed would be more likely to produce better output for most of its users.

\subsubsection{Study 2}
In Study 2, participants were only asked about \textit{either} personal trust \textit{or} general trust. The purpose of this was to avoid a potential confounding factor of seeing the combination of two questions and deliberately being asked to reflect on both oneself as a user and users more general. Group A, who were asked only about personal trust, consisted of 307 participants, of which 304 fully completed the survey. Group B, who were asked only about general trust/reliance, consisted of 314 participants, of which 300 fully completed the survey.

\subsection{Data analysis}
For the research questions about whether the \textit{proportion} of people that chose a product in an anthropomorphic description (RQ1, RQ1a, and RQ2) is higher than a hypothetical 50/50 split, we used the Chi-squared goodness-of-fit test with the following hypotheses:

\begin{itemize}
    \item H0: People are equally likely to choose a product when it is described in anthropomorphized language as when it is described in de-anthropomorphized language.
    \item H1: People are \textit{not} equally likely to choose a product when it is described in anthropomorphized language as when it is described in de-anthropomorphized language.
\end{itemize}
In practice, this means we expect the proportion that chooses re-Commender to be the same no matter if they see the anthropomorphized or de-anthropomorphized re-Commender (but \textit{not} assuming that the preference for re-Commender would necessarily be 50\%). Because all participants have been asked to choose \textit{one} of the products, we calculate this with the Chi goodness of fit-test. 

    \label{tab:chigoodnessoffit}

For the research questions that investigate if there is an association between different groups of people and preference for anthropomorphized/de-anthropomorphized descriptions (RQ3), we used the Chi-squared test of independence, with variables of, e.g., gender, socio-economic status, or education level, on one axis and anthropomorphized/de-anthropomorphized as the variables on the other. For all statistical tests we adopt a confidence level of 95\%. For the open text-answers, we performed a \textit{thematic analysis} \cite{clarke2015thematic}. This process is further described in the appendix, section \ref{sec:qualitativeanalysis}.


\section{Results}

\subsection{RQ1: Are people more likely to trust products that are described in anthropomorphizing language than products which are \textit{not} described in anthropomorphizing language?}

The results of Study 1 and Study 2 per product pair are shown in Table \ref{tab:overallresults}.

\begin{table}[]
\scalebox{0.7}{
\begin{tabular}{|l|llllll|}
\hline
\rowcolor[HTML]{EFEFEF} 
\multicolumn{7}{|c|}{\textbf{Study 1}} \\ \hline
\multicolumn{7}{|c|}{\texttt{Personal trust} / \color[HTML]{7091b5} \texttt{General trust}} \\ \hline
 \textbf{Category} & & \textbf{Ant.} & \textbf{De-ant.} & \textbf{\% ant.} & \textbf{$\chi^2$} & $p$ \\ \hline

\multirow{4}{1.2cm}{\textbf{Cognizer \newline 
($p$=\texttt{\textbf{.001$\ast$}/\newline\color[HTML]{7091b5}.012$\ast$})}} & 
reC & \texttt{114/\color[HTML]{7091b5}110} & \texttt{98/\color[HTML]{7091b5}95} & \texttt{53.8/\color[HTML]{7091b5}53.7} & & \\ 
& IntelliT & \texttt{71/\color[HTML]{7091b5}74} & \texttt{40/\color[HTML]{7091b5}44} & \texttt{64/\color[HTML]{7091b5}62.7} & \texttt{6.84/\color[HTML]{7091b5}6.27} & \texttt{\textbf{.009$\ast$}/\color[HTML]{7091b5}\textbf{.012$\ast$}} \\ \cline{2-7}

& MonAI & \texttt{97/\color[HTML]{7091b5}102} & \texttt{80/\color[HTML]{7091b5}94} & \texttt{54.8/\color[HTML]{7091b5}52} & & \\ 
& Cameron & \texttt{81/\color[HTML]{7091b5}67} & \texttt{61/\color[HTML]{7091b5}56} & \texttt{57/\color[HTML]{7091b5}54.5} & \texttt{4.29/\color[HTML]{7091b5}1.13} & \texttt{\textbf{.04$\ast$}/\color[HTML]{7091b5}.29} \\ \hline

\multirow{4}{1.2cm}{Agency} 
& HaulIT & \texttt{94/\color[HTML]{7091b5}92} & \texttt{81/\color[HTML]{7091b5}76} & \texttt{53.7/\color[HTML]{7091b5}54.8} & & \\ 
& Commuter & \texttt{68/\color[HTML]{7091b5}73} & \texttt{71/\color[HTML]{7091b5}73} & \texttt{48.9/\color[HTML]{7091b5}50} & \texttt{0.32/\color[HTML]{7091b5}0.82} & \texttt{.57/\color[HTML]{7091b5}.37} \\ \cline{2-7}

& AquaS & \texttt{64/\color[HTML]{7091b5}74} & \texttt{77/\color[HTML]{7091b5}72} & \texttt{45.4/\color[HTML]{7091b5}50.7} & & \\ 
& AI Scan & \texttt{71/\color[HTML]{7091b5}76} & \texttt{102/\color[HTML]{7091b5}92} & \texttt{41/\color[HTML]{7091b5}45.2} & \texttt{6.17/\color[HTML]{7091b5}0.62} & \texttt{\textbf{.013$\ast$}/\color[HTML]{7091b5}.43} \\ \hline

\multirow{4}{1.2cm}{Biological metaphors} 
& Judy & \texttt{60/\color[HTML]{7091b5}62} & \texttt{89/\color[HTML]{7091b5}89} & \texttt{40.3/\color[HTML]{7091b5}41.1} & & \\ 
& JurisD & \texttt{88/\color[HTML]{7091b5}88} & \texttt{78/\color[HTML]{7091b5}76} & \texttt{53/\color[HTML]{7091b5}53.7} & \texttt{1.15/\color[HTML]{7091b5}0.71} & \texttt{.28/\color[HTML]{7091b5}.40} \\ \cline{2-7}

& MindH & \texttt{68/\color[HTML]{7091b5}69} & \texttt{78/\color[HTML]{7091b5}81} & \texttt{46.6/\color[HTML]{7091b5}46} & & \\ 
& DermAI & \texttt{85/\color[HTML]{7091b5}82} & \texttt{87/\color[HTML]{7091b5}86} & \texttt{49.4/\color[HTML]{7091b5}48.8} & \texttt{0.45/\color[HTML]{7091b5}0.81} & \texttt{.50/\color[HTML]{7091b5}.37} \\ \hline

\multirow{4}{1.2cm}{Commu-\newline nicator} 
& Lingua & \texttt{86/\color[HTML]{7091b5}74} & \texttt{83/\color[HTML]{7091b5}86} & \texttt{50.9/\color[HTML]{7091b5}46.3} & & \\ 
& MentorMe & \texttt{82/\color[HTML]{7091b5}79} & \texttt{70/\color[HTML]{7091b5}82} & \texttt{53.9/\color[HTML]{7091b5}49.1} & \texttt{0.70/\color[HTML]{7091b5}0.70} & \texttt{.40/\color[HTML]{7091b5}.40} \\ \cline{2-7}

& WardrobE & \texttt{30/\color[HTML]{7091b5}35} & \texttt{32/\color[HTML]{7091b5}27} & \texttt{48.4/\color[HTML]{7091b5}56.5} & & \\ 
& Shoppr & \texttt{133/\color[HTML]{7091b5}138} & \texttt{125/\color[HTML]{7091b5}120} & \texttt{51.6/\color[HTML]{7091b5}53.5} & \texttt{0.11/\color[HTML]{7091b5}2.11} & \texttt{.74/\color[HTML]{7091b5}.15} \\ \hline

\multicolumn{7}{|l|}{$\chi^2$  = \texttt{29.75/\color[HTML]{7091b5}23.23}; $N$ = \texttt{2544/\color[HTML]{7091b5}2544}; $p$ = \texttt{\textbf{.013$\ast$}/\color[HTML]{7091b5}.079}} \\ \hline

\end{tabular}
\quad
\begin{tabular}{|l|llllll|}
\hline
\rowcolor[HTML]{EFEFEF} 
\multicolumn{7}{|c|}{\textbf{Study 2}} \\ \hline
\multicolumn{7}{|c|}{\texttt{Personal trust} / \color[HTML]{7091b5} \texttt{General trust}} \\ \hline
 \textbf{Category} & & \textbf{Ant.} & \textbf{De-ant.} & \textbf{\% ant.} & \textbf{$\chi^2$} & $p$ \\ \hline

\multirow{4}{1.2cm}{Cognizer} 
& Lingua & \texttt{72/\color[HTML]{7091b5}77} & \texttt{80/\color[HTML]{7091b5}66} & \texttt{47.4/\color[HTML]{7091b5}53.8} & & \\ 
& MentorMe & \texttt{71/\color[HTML]{7091b5}87} & \texttt{82/\color[HTML]{7091b5}74} & \texttt{46.4/\color[HTML]{7091b5}54} & \texttt{0.03/\color[HTML]{7091b5}1.89} & \texttt{.87/\color[HTML]{7091b5}.17} \\ \cline{2-7}

& WardrobE & \texttt{33/\color[HTML]{7091b5}28} & \texttt{30/\color[HTML]{7091b5}27} & \texttt{52.4/\color[HTML]{7091b5}50.9} & & \\ 
& Shoppr & \texttt{128/\color[HTML]{7091b5}117} & \texttt{114/\color[HTML]{7091b5}131} & \texttt{52.9/\color[HTML]{7091b5}47.2} & \texttt{0.005/\color[HTML]{7091b5}0.56} & \texttt{.94/\color[HTML]{7091b5}.46} \\ \hline

\multirow{4}{1.2cm}{\textbf{Agency}\newline($p$=\texttt{\textbf{.015$\ast$}/\newline\color[HTML]{7091b5}.97})} 
& reC & \texttt{102/\color[HTML]{7091b5}94} & \texttt{88/\color[HTML]{7091b5}88} & \texttt{53.7/\color[HTML]{7091b5}51.6} & & \\ 
& IntelliT & \texttt{64/\color[HTML]{7091b5}63} & \texttt{52/\color[HTML]{7091b5}62} & \texttt{55.2/\color[HTML]{7091b5}50.4} & \texttt{2.21/\color[HTML]{7091b5}0.16} & \texttt{.14/\color[HTML]{7091b5}.69} \\ \cline{2-7}

& MonAI & \texttt{110/\color[HTML]{7091b5}85} & \texttt{89/\color[HTML]{7091b5}92} & \texttt{55.3/\color[HTML]{7091b5}48} & & \\ 
& Cameron & \texttt{60/\color[HTML]{7091b5}62} & \texttt{46/\color[HTML]{7091b5}63} & \texttt{56.6/\color[HTML]{7091b5}49.6} & \texttt{4.02/\color[HTML]{7091b5}0.21} & \texttt{\textbf{.045$\ast$}/\color[HTML]{7091b5}.65} \\ \hline

\multirow{4}{1.2cm}{Biological metaphors} 
& HaulIT & \texttt{92/\color[HTML]{7091b5}84} & \texttt{92/\color[HTML]{7091b5}93} & \texttt{50/\color[HTML]{7091b5}47.5} & & \\ 
& Commuter & \texttt{57/\color[HTML]{7091b5}64} & \texttt{64/\color[HTML]{7091b5}61} & \texttt{47.1/\color[HTML]{7091b5}51.2} & \texttt{0.24/\color[HTML]{7091b5}0.12} & \texttt{.62/\color[HTML]{7091b5}.73} \\ \cline{2-7}

& AquaS & \texttt{80/\color[HTML]{7091b5}80} & \texttt{69/\color[HTML]{7091b5}62} & \texttt{53.7/\color[HTML]{7091b5}56.3} & & \\ 
& AI Scan & \texttt{71/\color[HTML]{7091b5}79} & \texttt{85/\color[HTML]{7091b5}81} & \texttt{45.5/\color[HTML]{7091b5}49.4} & \texttt{2.04/\color[HTML]{7091b5}0.85} & \texttt{.15/\color[HTML]{7091b5}.36} \\ \hline

\multirow{4}{1.2cm}{Commu-\newline nicator}
& Judy & \texttt{68/\color[HTML]{7091b5}86} & \texttt{84/\color[HTML]{7091b5}76} & \texttt{44.7/\color[HTML]{7091b5}53.1} & & \\ 
& JurisD & \texttt{85/\color[HTML]{7091b5}68} & \texttt{68/\color[HTML]{7091b5}72} & \texttt{55.6/\color[HTML]{7091b5}48.6} & \texttt{0.003/\color[HTML]{7091b5}0.12} & \texttt{.96/\color[HTML]{7091b5}.73} \\ \cline{2-7}

& MindH & \texttt{80/\color[HTML]{7091b5}85} & \texttt{75/\color[HTML]{7091b5}64} & \texttt{51.6/\color[HTML]{7091b5}57} & & \\ 
& DermAI & \texttt{79/\color[HTML]{7091b5}75} & \texttt{71/\color[HTML]{7091b5}78} & \texttt{52.7/\color[HTML]{7091b5}49} & \texttt{0.03/\color[HTML]{7091b5}1.07} & \texttt{.96/\color[HTML]{7091b5}.30} \\ \hline

\multicolumn{7}{|l|}{$\chi^2$  = \texttt{1.57/\color[HTML]{7091b5}0.80}; $N$ = \texttt{2441/\color[HTML]{7091b5}2424}; $p$ = \texttt{2.1/\color[HTML]{7091b5}.37}} \\ \hline

\end{tabular}}

\vspace{2mm}
    \caption{Results per product in Study 1 and Study 2. We indicate the $\chi^2$-values per product pair (as compared to an equal distribution between the anthropomorphized/deanthropomorphized description of each product). The `\% pref. ant.'-column indicates if the preference leans towards anthropomorphization (>50\%) or towards de-anthropomorphization (<50\%). Statistically significant values are indicated with \textbf{bold font} and a $\ast$-symbol. This table also indicates statistically significant $\chi^2$-values in the categories \textit{Cognizer} and \textit{Agency} -- these results are elaborated in Tables \ref{tab:categories1}-\ref{tab:categories3} in the appendix, section \ref{sec:appendix_resultscategories}.}
    \label{tab:overallresults}
\end{table}

\subsubsection{Study 1, personal trust}
1292 choices were made of the anthropomorphized product description, and 1252 choices were made of de-anthropomorphized product descriptions. The Chi-squared goodness-of-fit test showed that the distribution of preferences for anthropomorphized descriptions  was consistent with the H0 distribution ($\chi^2$ = 0.63; df = 14; $p$ = .43), meaning \textbf{there was no statistically significant preference for neither anthropomorphized nor de-anthropomorphized descriptions overall}. A Chi-squared test of independence shows \textbf{a statistically significant association between the products as a variable and the anthropomorphized/de-anthropomorphized descriptions} ($\chi^2$  = 29.74; $N$ = 2544; $p$ = .01). 

Between individual product pairs, we see that the preference changes per product, sometimes leaning towards a preference for the anthropomorphized description, and sometimes leaning against it. The re-Commender/IntelliTrade (recmmender systems) pair shows a significant preference for the anthropomorphized descriptions for both products ($\chi^2$ = 6.84; df = 1; $p$ > .001). Similarly, in the MonAI Maker/Cameron (personal assistant) pair, there is a significant preference for the anthropomorphized descriptions for both products ($\chi^2$ = 4.29; df = 1; $p$ > .04). In the AquaSentinel/AI Scan Guards (drones) pair, there is a significant preference for the \textit{de-}anthropomorphized descriptions for both products ($\chi^2$ = 6.17; df = 1; $p$ > .01). Interestingly, for the Judy/JurisDecide (legal recommendations) pair, there was a significant preference for the de-anthropomorphized description of the Judy system, and a preference for the anthropomorphized version of the JurisDecide system 
($\chi^2$  = 5.12; $N$ = 315; $p$ = .02).

\subsubsection{Study 1, general trust} 1295 choices were made of the anthropomorphized product description, and 1249 choices were made of de-anthropomorphized product descriptions. The distribution of anthropomorphized descriptions was consistent with the H0 distribution ($\chi^2$  = 0.70; df = 14; $p$ = .40), meaning there was \textbf{no statistically significant preference for neither anthropomorphized nor de-anthropomorphized descriptions overall}. A Chi-squared test of independence shows no statistically significant association between the products as a variable and the anthropomorphized/de-anthropomorphized descriptions ($\chi^2$  = 23.23; $N$ = 2544; $p$ = .08).

Between individual product pairs, only the re-Commender/IntelliTrade pair shows a statistically significant preference for the anthropomorphized descriptions of both products ($\chi^2$  = 6.27; df = 1; $p$ = .01), and the Judy/JurisDecide pair reveals a preference for the anthropomorphized description of JurisDecide, but a preference for the \textit{de}-anthropomorphized description of Judy with the product as a dependent variable ($\chi^2$  = 5.00; $N$ = 315; $p$ = .02). 

\subsubsection{Study 2, personal trust}
1252 choices were made of the anthropomorphized product description, and 1189 choices were made of de-anthropomorphized product descriptions. The distribution of preferences for anthropomorphized descriptions was consistent with the H0 distribution ($\chi^2$ = 1.57; df = 14; $p$ = .56), meaning there was \textbf{no statistically significant preference for neither anthropomorphized nor de-anthropomorphized descriptions overall}. A Chi-squared test of independence shows no statistically significant association between the products as a variable and the anthropomorphized/de-anthropomorphized descriptions ($\chi^2$  = 13.52; $N$ = 2441; $p$ = .56).

Between individual product pairs, the only statistically significant result is a preference for the anthropomorphized descriptions of both the MonAI/Cameron products (personal assistant) pair ($\chi^2$ = 4.02; df = 1; $p$ = .04).

\subsubsection{Study 2, general trust} 1234 choices were made of the anthropomorphized product descriptions, and 1190 choices were made of de-anthropomorphized product descriptions. The distribution was consistent with the H0 distribution ($\chi^2$ = 0.80; df = 14; $p$ = .37), meaning there was \textbf{no statistically significant preference for neither anthropomorphized nor de-anthropomorphized descriptions overall}. A Chi-squared test of independence shows no statistically significant association between the products as a variable and the anthropomorphized/de-anthropomorphized descriptions ($\chi^2$  = 8.99; $N$ = 2424; $p$ = .88).

Within the individual product pairs, we see no statistically significant preferences for neither anthropomorphized nor de-anthropomorphized descriptions.

\subsubsection{Aggregate results  (Study 1 + Study 2), personal trust} 
Across both studies, 2544 choices were made of the anthropomorphized product description, and 2441 choices were made of de-anthropomorphized product descriptions. The general distribution does not differ significantly from the null hypothesis, meaning we find \textbf{no statistically significant preference for neither anthropomorphized nor de-anthropomorphized product descriptions overall} ($\chi^2$ = 2.13; df = 14; $p$ = .14). The Chi-squared statistic for the accumulated numbers shows \textbf{a significant association between the product type as a variable and preference for either anthropomorphized or de-anthropomorphized description} ($\chi^2$  = 34.06; $n$ = 4985; $p$ = .003). Between individual product pairs, we see a significant preference for the re-Commender/IntelliTrade ($\chi^2$ = 8.47; df = 1; $p$ = .004) and MonAI/Cameron ($\chi^2$ = 8.31; df = 1; $p$ = .004) pairs, and a preference for the de-anthropomorphized description of Judy, but for the anthropomorphized description of JurisDecide with the product as a significant dependent variable ($\chi^2$  = 8.50; $n$ = 620; $p$ = .003). 

\subsubsection{Aggregate results  (Study 1 + Study 2), general trust} 
2529 choices were made of the anthropomorphized product description, and 2439 choices were made of de-anthropomorphized product descriptions. The distribution is consistent with the H0 distribution ($\chi^2$ = 1.63; df = 14; $p$ = .09), meaning there was \textbf{no statistically significant preference for neither anthropomorphized nor de-anthropomorphized descriptions overall}. The Chi-squared test shows no significant association between the variable product type and preference for neither anthropomorphized nor de-anthropomorphized description ($\chi^2$  = 9.02; $n$ = 4968; $p$ = .20). The only product pair that shows a signficant difference from the H0 distribution is the re-Commender/IntelliTrade pair, where there is a preference for the anthropomorphized description of both products ($\chi^2$ = 4.29; df = 1; $p$ = .04).

\subsection{RQ1a: Are people more likely to trust anthropomorphized products for themselves (personal trust) as a user than for the general population (general trust)?} 

A Chi-squared test with anthropomorphization/de-anthropomorphization as the first variable and personal vs. general trust as the second variable shows \textbf{no significant relationship between the variables personal and general trust and preference for anthropomorphized/de-anthropomorphized descriptions}, neither in Study 1 ($\chi^2$  = 0.007; $N$ = 2544; $p$ = .93), nor in Study 2 ($\chi^2$  = 0.16; $n$ = 4841; $p$ = .68), nor in aggregate results ($\chi^2$  = 0.07; $n$ = 4865; $p$ = .79).

\subsection{RQ2: Are people more likely to trust products when the products are described in different \textit{kinds} of anthropomorphizing language?}

\paragraph{Personal trust}
The choices of anthropomorphized/de-anthropomorphized descriptions per category are shown in the appendix, section \ref{sec:antcategories}, Tables \ref{tab:categories1}, \ref{tab:categories2}, and \ref{tab:categories3}. 

For Study 1, a Chi-squared test of independence shows \textbf{a statistically significant association between the categories as a variable and the preference for anthropomorphized/de-anthropomorphized descriptions} ($\chi^2$  = 14.41; $n$ = 2544; $p$ = .002). The \textit{Properties of a cognizer}-category is the only category with a distribution that differs significantly from the H0 distribution ($\chi^2$ = 10.99; df = 1; $p$ < .001). For Study 2, a Chi-squared test of independence shows \textbf{no statistically significant association between the categories and the anthropomorphized/de-anthropomorphized descriptions} ($\chi^2$  = 4.96; $n$ = 2441; $p$ = .17), but it did show a significant preference for the anthropomorphized descriptions in the \textit{Agency}-category ($\chi^2$ = 5.89; df = 1; $p$ < .01). It is worth noting that the product pairs in the \textit{Cognizer}-category in Study 1 (recommender systems and personal assistants) were the same products as had been assigned the \textit{Agency}-category in Study 2 (as shown in Table \ref{tab:overallresults}). Hence, those specific products or product categories may be especially prone to preference in anthropomorphized descriptions (no matter the type of anthropomorphizing language).

If we aggregate the numbers from both studies, there is \textbf{no statistically significant association between the language categories and the preference for anthropomorphized/de-anthropomorphized descriptions} ($\chi^2$  = 6.16; $n$ = 2441; $p$ = .10), but there is \textbf{a significant preference for the anthropomorphized descriptions in the \textit{Cognizer}-category} alone ($\chi^2$ = 5.37; df = 1; $p$ = .02). This is a spillover effect: in Study 1, the preference for anthropomorphized descriptions in the \textit{Cognizer}-category is so strong (56.5\% and 55\% for personal and general trust, respectively), that despite a very slight negative preference in personal trust in Study 2 (49.8\%) and a weaker preference in general trust (50.9\%) the preference carries over.

\paragraph{General trust} In Study 1, there is \textbf{no statistically significant association between the categories and the preference for anthropomorphized/de-anthropomorphized descriptions} ($\chi^2$  = 7.26; $n$ = 2544; $p$ = .06), but there is a significant preference for the anthropomorphized descriptions in the \textit{Cognizer}-category ($\chi^2$ = 6.38; df = 1; $p$ = .01). In Study 2, there was \textbf{no statistically significant association between the categories and the preference for anthropomorphized/de-anthropomorphized descriptions} ($\chi^2$  = 0.52; $n$ = 2424; $p$ = .91), and no significant difference from the H0 distribution in either of the categories. Aggregating the numbers, a Chi-squared test of independence shows \textbf{no association between the categories and the preference for anthropomorphized/de-anthropomorphized descriptions} ($\chi^2$  = 4.21; $n$ = 4968; $p$ = .24), but there is \textbf{a significant association in the \textit{Cognizer}-category} ($\chi^2$ = 4.50; df = 1; $p$ = .03).

\subsection{RQ3: Are different groups of people more likely to trust products when the products are described in anthropomorphizing language?}

The values from the statistical tests are shown in Table \ref{variablesaggregate}. We refer to the appendix, Tables \ref{tab:gender1}-\ref{tab:comp3} for detailed results per study. We highlight that we focus on Chi-squared statistics for \textit{the entire variable}, i.e., the association between a variable and proportion of choices of anthropomorphized/de-anthromoporphized descriptions. There can be significant preferences within each subgroup (e.g., \textit{female} vs. \textit{male}, but due to space restrictions we only discuss the variables where the entire chi-squared statistic is significant)

\begin{table}[]
\small
\renewcommand{\arraystretch}{1.2}
\begin{tabular}{|l|m{1.5cm}|m{1.5cm}|m{1.5cm}|m{1.5cm}|m{1.5cm}|m{1.5cm}|}
\hline
\textbf{Variable} & \multicolumn{3}{c}{\textbf{Personal Trust}} & \multicolumn{3}{|c|}{\textbf{General Trust}} \\ \hline

 & \textbf{Study 1} & \textbf{Study 2} & \textbf{Aggregate} & \textbf{Study 1} & \textbf{Study 2} & \textbf{Aggregate} \\ \hline

\renewcommand{\arraystretch}{1.2}
\textbf{Gender} 
& \makecell[l]{$\chi^2$  = 4.00 \\ $N$ = 2505 \\ $p$ = .13} 
& \makecell[l]{$\chi^2$  = 0.5 \\ $N$ = 2416 \\ $p$ = .48} 
& \makecell[l]{$\chi^2$  = 1.00 \\ $N$ = 4921 \\ $p$ = .60} 
& \makecell[l]{$\chi^2$  = 4.44 \\ $N$ = 2505 \\ $p$ = .10} 
& \makecell[l]{$\chi^2$  = 0.12 \\ $N$ = 2400 \\ $p$ = .94} 
& \makecell[l]{$\chi^2$  = 2.43 \\ $N$ = 4905 \\ $p$ = .30} \\ \hline

\textbf{Age} 
& \makecell[l]{$\chi^2$  = 12.99 \\ $N$ = 2512 \\ $p$ = .22} 
& \makecell[l]{\textbf{$\chi^2$  = 21.06} \\ \textbf{$N$ = 2432} \\ \textbf{$p$ = .02$\ast$}}
& \makecell[l]{\textbf{$\chi^2$  = 18.45} \\ \textbf{$N$ = 4944} \\ \textbf{$p$ = .048$\ast$}} 
& \makecell[l]{$\chi^2$  = 10.07 \\ $N$ = 2512 \\ $p$ = .43} 
& \makecell[l]{$\chi^2$  = 10.50 \\ $N$ = 2400 \\ $p$ = .40} 
& \makecell[l]{$\chi^2$  = 14.51 \\ $N$ = 4912 \\ $p$ = .20} \\ \hline

\textbf{Socio-economic status} 
& \makecell[l]{$\chi^2$  = 6.09 \\ $N$ = 2493 \\ $p$ = .19} 
& \makecell[l]{$\chi^2$  = 2.23 \\ $N$ = 2424 \\ $p$ = .69} 
& \makecell[l]{$\chi^2$  = 2.64 \\ $N$ = 4917 \\ $p$ = .62} 
& \makecell[l]{$\chi^2$  = 3.40 \\ $N$ = 2493 \\ $p$ = .49} 
& \makecell[l]{$\chi^2$  = 6.46 \\ $N$ = 2392 \\ $p$ = .17} 
& \makecell[l]{$\chi^2$  = 7.08 \\ $N$ = 4885 \\ $p$ = .13} \\ \hline

\textbf{Level of education} 
& \makecell[l]{$\chi^2$  = 7.63 \\ $N$ = 2513 \\ $p$ = .11} 
& \makecell[l]{$\chi^2$  = 4.17 \\ $N$ = 2432 \\ $p$ = .24} 
& \makecell[l]{$\chi^2$  = 6.99 \\ $N$ = 4888 \\ $p$ = .07} 
& \makecell[l]{$\chi^2$  = 4.87 \\ $N$ = 2513 \\ $p$ = .30} 
& \makecell[l]{$\chi^2$  = 1.96 \\ $N$ = 2400 \\ $p$ = .74} 
& \makecell[l]{$\chi^2$  = 1.78 \\ $N$ = 4816 \\ $p$ = .62} \\ \hline

\textbf{Level of computer knowledge} 
&\makecell[l]{$\chi^2$  = 1.49 \\ $N$ = 2504 \\ $p$ = .68} 
& \makecell[l]{$\chi^2$  = 1.85 \\ $N$ = 2432 \\ $p$ = .60} 
& \makecell[l]{$\chi^2$  = 2.30 \\ $N$ = 4928 \\ $p$ = .51} 
& \makecell[l]{$\chi^2$  = 0.35 \\ $N$ = 2504 \\ $p$ = .95} 
& \makecell[l]{$\chi^2$  = 0.47 \\ $N$ = 2400 \\ $p$ = .93} 
& \makecell[l]{$\chi^2$  = 0.73 \\ $N$ = 4896 \\ $p$ = .87} \\ \hline

\end{tabular}
\caption{Results of Chi-squared tests for each variable. Statistically significant results are marked in \textbf{bold font} and with a $\ast$. The detailed results are provided in the appendix, Tables \ref{tab:gender1}-\ref{tab:comp3}.} 
\label{variablesaggregate}
\end{table}

\subsubsection{Self-described gender}

The proportion of choices of anthropomorphized/de-anthropomorphized product descriptions \textbf{did not differ significantly by gender} in either study, neither in personal trust, nor in general trust.

\subsubsection{Age}

In Study 1, a Chi-test of independence showed no significant relationship between the two variables age and proportion of choices of anthropomorphized/de-anthropomorphized descriptions. In Study 2, the same test showed a significant relationship between the variables, and this repeated for the aggregate results, meaning there was \textbf{an overall significant association between different age groups and their preference for anthropomorphized or de-anthropomorphized product descriptions in personal trust}.

Looking closer at the age groups individually, only the 61-65 year group shows a strong, statistically significant preference for anthropomorphized descriptions ($\chi^2$ = 14.70; df = 1; $p$ < .001). In some of the age groups, $n$ is too small to draw meaningful conclusions within different categories, but we highlight a significant preference for the anthropomorphized descriptions for groups 31-35 and 51-55 in the \textit{Cognizer}-category ($\chi^2$ = 4.40; df = 1; $p$ = .04 and $\chi^2$ = 3.88; df = 1; $p$ = .05, respectively), in the 36-40 age group, there was a significant preference for the anthropomorphized descriptions in the \textit{Communicator}-category ($\chi^2$ = 6.92; df = 1; $p$ = .01), and in the 41-45 age group, there was a strong preference for the \textit{de}-anthropomorphized descriptions in the \textit{Biological metaphors}-category ($\chi^2$ = 3.97; df = 1; $p$ = .05). \textbf{No statistically significant association between age and preference for anthropomorphized/de-anthropomorphized product descriptions could be found in general trust} in either study.

\subsubsection{Socio-economic status}

A chi-squared test showed that the proportion of choices of anthropomorphized/de-anthropomorphized product descriptions \textbf{did not differ significantly by socio-economic status in either study}, neither in personal trust, nor in general trust.

\subsubsection{Level of education}

\textbf{No significant association was found between level of education and preference for anthropomorphized or de-anthropomorphized descriptions} in either study, neither for personal trust, nor for general trust.

\subsubsection{Level of computer knowledge}

The proportion of choices of anthropomorphized/de-anthropomorphized product descriptions \textbf{did not differ significantly by level of computer knowledge} in either study, neither in personal trust, nor in general trust.

\vspace{-0.5em}
\section{Discussion}
\vspace{-0.5em}
The qualitative responses from the surveys circumstantiate and add detail to the quantitative results. Because the open answers were optional, we do not attempt to quantify their importance or weight in any way, and that would be meaningless: since some product pairs had 30 elaborations, while some had maybe 100, some insights might be unfairly under- or over-represented. We use the open answers to shed light on a complex topic and study, and to provide insights that hopefully lead to fair and purposeful future investigation in the subject.

\vspace{-0.5em}

\subsection{Observation 1: Overall, people are no more likely to choose anthropomorphized descriptions of products over de-anthropomorphized descriptions of probabilistic automation products.} \label{sec:obs1}

Across categories, we do not see a clear preference for anthropomorphized descriptions of products over de-anthropomorphized descriptions of products. This is a conclusion that come with numerous addenda, the most important one being \textit{``it depends''}\,---\,within some product descriptions there was a significant preference for the anthropomorphized description, and for some systems there was a clear preference for the \textit{de}-anthropomorphized description. The preference proportions changed between the two studies, after anthropomorphization categories were swapped. This points to the conclusion that both product genre and type of anthropomorphization influence how people immediately perceive a product based on its description. A few participants even highlighted linguistic differences in product descriptions as motivating their choice, albeit using different words than anthropomorphization: \textit{``Option B provides a more engaging and descriptive presentation ''} (Study 1, de-ant. AquaSentinel\footnote{The parentheses after quotes are in the form [Study \#, anthropomorphized/de-anthropomorphized description, product], in this case indicating that the participant was part of Study 1, the participant chose the de-anthropomorphized description of AquaSentinel (therefore comparing it to the anthropomorphized AI Scan Guards).}). We find the following main themes or clusters when looking for how participants motivate their rationale:

\vspace{-0.5em}
\paragraph{Lesser of two evils-motivation}
A prevalent theme in the open text answers is that the participant has chosen ``the lesser of two evils''; meaning they are expressing deep skepticism of both products, but was forced, through the survey design, to choose one. In this case, the motivation appears to be identifying which product has \textit{lower stakes}, or less impact if the system somehow fails: \textit{``Lower stakes - only deals with hobbies/past times as opposed to finances''} (Study 1, ant. re-Commender), and \textit{``I would trust AI more to transport goods than people''} (Study 1, de-ant. HaulIt).

\vspace{-0.5em}
\paragraph{People attempt to evaluate shortcomings and strengths of using probabilistic automation for the particular context}
A lot of responses express that probabilistic automation is more appropriate for some tasks than for others. For instance, most responses in favor of the MonAI system in favor of the Cameron system highlight that \textit{``Computers are better with numbers than texts. I would trust more an app with numbers than one who manage texts.''} (Study 1, ant. MonAI Maker). However, many of these assumptions are exactly that; \textit{assumptions} of the system's functionality: \textit{``It will be more correct because it works with photos for comparison, so the chance of error is smaller''} (Study 2, de-ant. DermAI Scan). This is hardly an objective truth, and broad assumptions like this emphasize the importance of conveying accurate expectations of the system's functionality, because people are prone to form beliefs even based on short descriptions. 
The logic appears to be, of course, that the perceived benefits should outweigh the potential risks.

\vspace{-0.5em}
\paragraph{Human favoritism}
A common theme in the responses was \textit{human favoritism}, perceiving an output as higher quality if a human expert has been involved in the process of creating it \cite{zhang2023human}. This was visible as expressions of preference for the products where a human was assumed in control of the probabilistic automation product, even when this was not actually described in the product pitch, e.g., \textit{``There is both a person driving it and an AI in it''} (Study 1, de-ant. Commuter). The potential of biased probabilistic automation training data was mentioned many times as rationale for distrusting the system (e.g., \textit{``I dislike the idea of AI in the justice system when it is prone to making up information. How do we know that Judy would be free from bias?''} (Study 2, de-ant. JurisDecide). Human favoritism is an interesting notion, in that it could potentially introduce issues of over-reliance on human judgments and under-estimation of human bias and error-proneness.

Overall, our results show that people do not unequivocally trust technology just because it is linguistically anthropomorphized. People are critical about use context, risks, impacts, and human involvement, and although we confirm earlier research that demonstrate some influence of anthropomorphization on attitude (e.g., \cite{langer2022look, kim2023communicating}), there is not a binary or simple relationship between anthropomorphization and trust. 

\paragraph{De-anthropomorphization carries a risk of misunderstandings} 
A very interesting finding was that a few users simply did not understand the de-anthropomorphized (but more technically accurate) descriptions as examples of probabilistic automation products, e.g. \textit{``I'm not sure I would entirely trust Cameron not to miss any important/urgent emails. However when it came to my data I'd trust it more than any AI.''} (Study 1, de-ant. Cameron\footnote{In the de-antropomorphized version, Cameron was described as ``powered by automatic pattern matching'' instead of ``powered by artificial intelligence''.}). This person appears to express a general aversion to the concept ``AI'', and has not picked up that ``automatic pattern matching'' is actually the same as ``AI''. The de-anthropomorphized description leads to a misunderstanding. Other examples are \textit{``I prefer this to AI''} (Study 2, ant. MindHealth) and \textit{``this one doesn't use neural networks so it's most likely to be more accurate''} (Study 1, de-ant. JurisDecide\footnote{``Neural networks'' was swapped for ``weighted networks'' in the de-anthropomorphized description}). This is a significant risk that we need to consider when describing probabilistic automation systems: how do we balance the advantages of using language and metaphors that people are familiar with, with the risks of those analogies and metaphors leading to incorrect assumptions?

\vspace{-0.8em}

\subsection{Observation 1a: Across the two studies, people are no more likely to trust anthropomorphized product descriptions when imagining themselves as a user than to trust them for the general population}
\label{sec:discussionpersonalvsgeneral}

\vspace{-0.2em}

In both studies, several trends in preference under personal trust were not present when asked about general trust. This was the case both in Study 1, where participants were asked about both personal and general trust per product, and in Study 2, where each participant was only asked about either personal or general trust. For Study 1, we suspected there could be an ordering effect of the survey; the first question might elicit an immediate response, and the immediate invitation to reflect again on the product in relation to general trust could urge the participant to feel they should choose something different for the second option. This, however, does not explain the differences in Study 2, where the participant groups were different for the personal trust and for the general trust questions. 

In fact, we see for Study 2 that preferences (see Table \ref{tab:overallresults}) lean in different directions for several product pairs, and overall for the different categories (\textit{Cognizer}, \textit{Agency}, and \textit{Biological metaphors} all elicit different preferences between personal and general trust in Study 2). The differences are small, however (e.g., 48.9\% preference for anthropomorphized descriptions for personal trust vs. 50.8\% preference for anthropomorphized descriptions in general trust for the \textit{Cognizer}-category), and none of them are statistically significant in the overall comparison, except for the \textit{Agency}-category, which elicited 55\% and 49.9\% preference for the anthropomorphized descriptions in personal and general trust, respectively. We could not identify any obvious differences in the qualitative responses between participants' rationale for choosing products for themselves and evaluating their output in general.

\subsection{Observation 2: The type of product or system in combination with different kinds of anthropomorphizing language appears to exert a greater influence on trust than anthropomorphizing language alone.}

Since we saw a statistically significant association between \textit{product type} as a variable and preference for anthropomorphized/de-anthropomorphized descriptions in personal trust in Study 1, we decided to change the categories of anthropomorphizing language between products and conduct the second study to explore this potentially confounding variable. The fact that the products in the recommender systems and personal assistants resulted in a preference for anthropomorphized descriptions in the \textit{Cognizer} category in Study 1, \textit{and} in the \textit{Agency} category in Study 2 (at least in personal trust), indicates that certain products or systems might be more sensitive to anthropomorphized language than others. Interestingly, this goes in both `directions': the `Judy' and the `AI Scan Guards' systems were generally more trusted in the \textit{de-anthropomorphized} descriptions. We note, that these systems were both in the \textit{Biological metaphors} category in Study 1 and Study 2, respectively --- we hypothesize that this category of language may yield particularly contrived analogies which approach the uncanny valley \cite{mori2012uncanny} and, consequently, mistrust. This, however, does not explain the general preference for \textit{anthropomorphized} descriptions of `JurisDecide' and `AquaSentinel' --- the two products that `Judy' and `AI Scan Guards' were compared to, and which were in the same language categories (Biological metaphors).

Our findings advocate for a nuanced conclusion that the individual product or system is an important variable for people's preferences and attribution of trustworthiness. Some products might be more susceptible to anthropomorphization of one type, and certain types of anthropomorphization might highlight or obfuscate specific qualities in specific system genres. Our studies thus support the findings of \cite{langer2022look}.




\subsection{Observation 3: Age is the only variable that seems to have a dependent association with preferences for anthropomorphized/de-anthropomorphized product descriptions.}

When dividing participants into subgroups by age, some patterns emerge per category as well as overall. Interestingly, we see a strong preference for anthropomorphized descriptions in the 61-65 group, and a strong preference for \textit{de}-anthropomorphized descriptions in the 66+ group. The subgroups are small, however, (26 participants total for the 61-65 group, and 37 for the 66+ group), so we refrain from making general conclusions on the basis of this study. The groups 31-35 and 36-40 compose a larger proportion of participants, and these groups both show a strong preference for anthropomorphized descriptions, particularly in the \textit{Cognizer}-category. When looking at the open answers, these age groups do not seem to provide different rationales from other age groups; they (also) highlight factors such as \textbf{personal usefulness} (\textit{``I can grocery shop weekly [...] but I am always surprised by the fact that ALL my basics become [worn] out at the same time''}(Study 1, ant. WardrobEase)), \textbf{privacy} (\textit{``I would never use my voice online''} (Study 2, de-ant. DermAI Scan)) \textbf{risk of failure} (\textit{``I trust AI Scan Guards to give better output, due to its systems having less of a chance to be disrupted by enemy counter electronics warfare''} (Study 2, ant. AI Scan Guards)) and \textbf{impact in case of failure} (\textit{``[AI] dealing with the jury can skew what their outcomes would be.''} (Study 2, de-ant. JurisDecide)) as the main motivations behind their choices. One hypothesis to explain these differences across age groups is that there could be age-related factors influencing computing literacy for different groups . A recent survey has indicated a generation gap in probabilistic automation-acceptance \cite{pcmagGenerationGap}, 
and potentially, using more familiar language to describe such systems (playing on anthropomorphizing metaphors and analogies) may make the systems more appealing to these groups.


\section{Limitations}
We acknowledge study only explores a small part of the overarching question ``What are the effects of anthropomorphization of probabilistic automation systems?'' This question could be explored in many ways that are likely to provide other results. Some of the most important limitations to the approach used in this study are listed below:

\textbf{Contrived study setup rather than organic choice.} Any controlled experiment can impose confounding factors. ``Trust'' based on momentary, immediate choices, rather than long-term, more organic exposure to descriptions of a system. Conversely, one could argue that based on the qualitative answers, participants have relied heavily on their existing knowledge about probabilistic automation systems, so we are not exposing them to completely novel technology descriptions. Participants were also asked to choose based on only a short description and no examples of the system's output. We do not believe this was a confounding factor for the results, but it could mean that the results will not generalize to contexts where more information is given.

\textbf{Contrived language.} To emphasize the anthropomorphic language as a variable, we have loaded a lot of `anthropomorphisms' into very little text. A few participants highlighted linguistic or semantic features of the descriptions as determining factors for their choice (see section \ref{sec:obs1}), so it is possible that this would have impacted the results to some degree. We have tried to mitigate this factor by creating descriptions that are directly comparable to actual products found ``in the wild''.

\textbf{Not all categories were tested on all products.} We only swapped the categories between two different products. Ideally, we could have tried all categories of anthropomorphization on all product types, however, this would have required an untenable amount of different studies (and no page restrictions). The results provide enough insight for us to conclude that the matter is not straightforward, and that further investigation is needed.

\textbf{Order effects bias.} In the survey, product pairs were always presented in the same order, which could could induce order effects bias. This should not have any effect on the primary variable (anthropomorphized versus de-anthropomorphized), as these choices were always randomized. 

\section{Conclusions}
In this paper, we explored an overall question of the influence of anthropomorphized short descriptions of probabilistic automation systems on trust. 
We made three observations based on the results: 1. Across both studies, people were no more likely to prefer anthropomorphized products over de-anthropomorphized products. 2. The product type in combination with anthropomorphizing language appears to exert higher influence on trust than anthropomorphizing language alone, and 3. Age was the only variable (of those measured) which had a statistically significant association with preference for anthropomorphized vs. de-anthropomorphized products.

Our results show that anthropomorphized descriptions of systems do not automatically lead to favoritism or increased trust. It appears to depend on product category and type of anthropomorphization, as well as the reader of the text. We highlight that this was an exploratory study which hopefully provides inspiration for further investigation by other researchers. We hope that the results are useful to those who write about probabilistic automation systems, whether they be scholars, policy makers, or journalists. Our future work will include further exploration of empirically founded taxonomies of anthropomorphization, as well as more detailed studies of the risks of ``trust'', investigating different impact of anthropomorphized descriptions of probabilistic automation systems.

\section{Impact Statement} \label{sec:impact}
In designing our online survey we adhered to the ethical guidelines in HCI methodology \cite{bruckman2014research} to ensure participant anonymity and data privacy. Participants were recruited via the Prolific platform, and compensated for their participation. To ensure that we reached a representative group we created pre-screening
criteria such as country of residence, self-assessed socio-economic status, ethnicity, and geographic location. We did not collect any identifiable information and all the survey responses were stored temporarily on a secure server. To avoid confusion about the fictitious products, we added a statement at the end of the survey asserting that all products are 100\% imagined, although some of them have been loosely based on existing products or services. We also stated that the goal of the research was to investigate whether the description of the product influenced the way its trustworthiness and functionality is perceived, as well as contact info for the lead author.



\subsection{Positionality statement for the study authors.}
The expertise and lived experiences of our research team were an important part of the judgments and discussions in our analysis. We present our research team positionality according to the guidelines proposed by \citeauthor{liang2021embracing}.

The first author has a background in digital design and positions themselves as an enthusiast of (mixed methods) research methodology. Their research career has focused on understanding how people interact with technology, and how technology impacts human cognition. Their background shapes the work by increasing their attention to qualitative data as a primary resource for understanding quantitative results.   

The second author positions themselves primarily as an activist for better and more inclusive AI education. 
They worked for more than eight years on hands-on STEAM education in different communities worldwide as part of the organization they created called [\textit{Anonymized}].  In the past four years, they have led multiple co-design sessions with families focused on AI literacy and created \textit{[Anonymized]}, one of the first platforms for AI education, which is free and open-source. This experience influenced their focus on critical understanding and use of probabilistic automation systems and informed their understanding of how the perception of technology can shape people's trust and use of it.

The third author is a natural language processing scientist with a background in low-resource NLP and the digital documentation of resources for low-resource language communities. Their work also encompasses the field of human-computer interaction and the intersection between NLP and psycholinguistics. Their previous work in human-computer interaction and AI provides insights into how users perceive and interact with technology, contributing to a deeper understanding of trust dynamics in AI systems.

The fourth author is a computational linguist, with expertise in syntax, semantics and sociolinguistics. They have long worked at the intersection of linguistics and natural language processing, specifically on how linguistic knowledge can inform the development and study of language technology. They have been doing public scholarship around the way that probabilistic automation technologies are sold and perceived and advocating for more accurate and less aspirational descriptions of this technology.

We acknowledge that while our study addresses a timely question of how people's trust in automation driven systems can be influenced by different forms of anthropomorphism it could also lead to a potential dual use. For example, bad actors could use our findings to elicit unearned trust from people, in particular by describing technical systems functionality in cognitive terms and by emphasizing their ``intelligence''. Bad actors could also use the observations from our study to target specific age groups that seem to be more susceptible to trust systems with anthropomorphized descriptions.

\begin{acks}
Thank you to Sasha Luccioni, Leon Derczynski, and Alexander Koller for the discussions that helped frame this study. This research was supported by the VILLUM Foundation, grant 37176 (ATTiKA: Adaptive Tools for Technical Knowledge Acquisition). 
\end{acks}

\bibliographystyle{ACM-Reference-Format}
\bibliography{paper}

\appendix

\clearpage

\section{Survey screenshot}
\label{sec:surveyscreenshot}

To align understandings of trust, the survey was introduced with the following text: 

``On the following pages, you will be introduced to a series of technical systems. We ask you to evaluate these systems along the following two dimensions: 

\textbf{How likely you would be to trust the system as a user.}
We ask you to think about how likely you would be to trust using this system for your own purposes, assuming you would like to use the service it would provide.
    
\textbf{Which systems you think would generally give better output for its users.}
Here, we ask you to think about the general quality of output that this system would produce. So, even if the system didn’t provide a service relevant to you, would it be a good system for its users?
    
Under each question you can provide further information about your rationale behind your choice – if you wish to do so.''

\begin{figure} [h!]
    \centering
    \includegraphics[width=11cm]{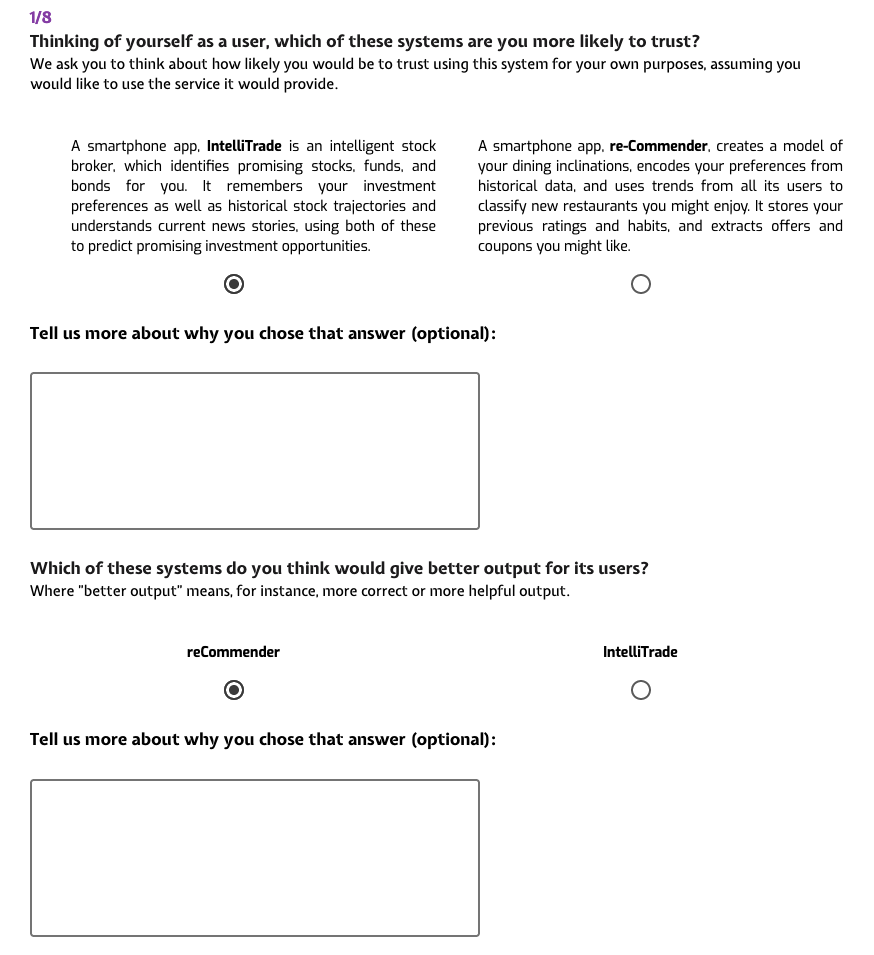}
    \caption{A screenshot of one page of the survey as it was presented to participants. Each product pair was presented on a separate page together with the open text option to elaborate.}
    \label{fig:surveyscreenshot}
\end{figure}

\section{Product descriptions}
In each product descriptions, instances of anthropomorphization (4-5 per product) are highlighted to allow for easier comparison to the de-anthropomorphized version. Each of the anthropomorphic short pitches was written to fit in its respective category, and each of the short pitches included 4-5 ``instances'' of the anthropomorphic category. For each de-anthropomorpized description of the product, the five instances of anthropomorphic language were re-written so they did not reflect the specific category of anthropomorphization, but the rest of the short pitch could include examples of the other categories of anthropomorphization\,---\,thus, isolating each anthropomorphization category as the independent variable. We were not strict about avoiding other categories of anthropomorphic language (especially the category of agency) in the pitches. However, we also did not de-anthropomorphize language outside the target anthropomoprhic language type in the corresponding de-anthropomorphized product description. For example, in Study 1, MonAI Maker is described as {\it identifying ways to save money}, a cognizer description and de-anthropomorphized as {\it providing suggestions} instead. This is still agentive language.

\begin{table}[h!]
\small
    \centering
    \renewcommand{\arraystretch}{1.4}%
    \scalebox{0.8}{
    \begin{tabular}{m{0.6cm}|m{4.1cm}|m{4.1cm}|m{4.1cm}|m{4.1cm}}
         \textbf{Cat.} & \textbf{Anthropomorphized} & \textbf{De-anthropomorphized} & \textbf{Anthropomorphized} & \textbf{De-anthropomorphized}  \\ \hline
         \rotatebox{90}{\textbf{Properties of a \textit{cognizer}}} 
         & A smartphone app, \textbf{re-Commender}, \hl{understands} your dining preferences, \hl{knows} your preferences from historical data, and uses trends from all its users to \hl{predict} new restaurants you might enjoy. It \hl{remembers} your previous ratings and habits, and \hl{figures out} offers and coupons you might like. & A smartphone app, \textbf{re-Commender}, \hl{creates a model of} your dining inclinations, \hl{encodes} your preferences from historical data, and uses trends from all its users to \hl{classify} new restaurants you might enjoy. It \hl{stores} your previous ratings and habits, and \hl{extracts} offers and coupons you might like.
         & A smartphone app, \textbf{IntelliTrade} is an \hl{intelligent} stock broker, which \hl{identifies} promising stocks, funds, and bonds for you. It \hl{remembers} your investment preferences as well as historical stock trajectories and \hl{understands} current news stories, using both of these to \hl{predict} promising investment opportunities. 
         & A smartphone app, \textbf{IntelliTrade} is an \hl{automated} stock broker, which \hl{makes calculations about} promising stocks, funds, and bonds for you. It \hl{encodes} your investment preferences as well as historical stock trajectories and \hl{processes} current news stories, using both of these to \hl{classify} promising investment opportunities.
         \\ \hline
        \rotatebox{90}{\textbf{Properties of a \textit{cognizer}}}
        & \textbf{MonAIMaker} is an \hl{intelligent} app that helps you plan your personal finances. It \hl{learns} what you are likely to spend money on by \hl{recognizing} trends in your bank statements as well as your email correspondences. It uses these to \hl{identify} ways to save money, and \hl{remember} when you have bills and expenses due. 
        & \textbf{MonAIMaker} is an \hl{automatic pattern matching} app that helps you plan your personal finances. It \hl{classifies} what you are likely to spend money on by \hl{mapping} trends in your bank statements as well as your email correspondences. It uses these to \hl{provide suggestions for} ways to save money, and \hl{store information about} when you have bills and expenses due.
        & An app, \textbf{Cameron}, is powered by \hl{artificial intelligence} and \hl{machine learning} to help you organize and answer your emails. It \hl{interprets} text from your incoming emails, \hl{suggests} answers based on your writing style, and \hl{recognizes} tasks and deadlines to create automated to-do-lists for you. 
        & An app, \textbf{Cameron}, is powered by \hl{automatic pattern matching} and \hl{machine conditioning} to help you organize and answer your emails. It \hl{classifies} text from your incoming emails, \hl{synthesizes} answers based on your writing style, and \hl{assigns labels to} tasks and deadlines to create automated to-do-lists for you.
        \\ \hline
        \rotatebox{90}{\textbf{Agency}}
        & A \hl{self-driving} truck \textbf{HaulIT} \hl{handles} long haul freight 24/7 without rest stops, and it never gets tired or distracted. The truck is designed for both city and highway, meaning it always \hl{chooses} the most optimal route for speed and efficiency by \hl{analyzing} current and projected traffic conditions and \hl{self-managing} battery charging. 
        & A \hl{driverless} truck \textbf{HaulIT} \hl{is programmed to transport} long haul freight 24/7 without rest stops, and it never gets tired or distracted. The truck is designed for both city and highway, meaning it is always \hl{sent along} the most optimal route for speed and efficiency, \hl{based on statistical predictions about} current and projected traffic conditions as well as \hl{optimal} battery charging points. 
        & A sleeper bus, \textbf{Commuter}, \hl{drives} people from their home to a long-distance destination overnight. The bus \hl{avoids} other vehicles and obstacles on the road, and \hl{adapts} to the weather conditions to navigate safely. It \hl{monitors} traffic live and \hl{picks} the best and safest routes. & A sleeper bus, \textbf{Commuter}, \hl{is used to transport} people from their home to a long-distance destination overnight. The bus \hl{has algorithms for avoiding} other vehicles and obstacles on the road, and \hl{the algorithms are adjusted} to the weather conditions to navigate safely. It\hl{s systems are fed} live traffic \hl{data} for \hl{calculations} of the best and safest routes.
        \\ \hline
        \rotatebox{90}{\textbf{Agency}}
        & An AI and ML-powered drone, \textbf{AquaSentinel AI-MAR}, \hl{monitors} enemy seas. Armed with cutting-edge technology, it \hl{autonomously patrols} waterways, \hl{utilizing} advanced algorithms to swiftly \hl{detect} and \hl{analyze} potential threats in real-time. 
        & An AI and ML-powered drone, \textbf{AquaSentinel AI-MAR}, \hl{is programmed to monitor} enemy seas. Armed with cutting-edge technology, it \hl{is positioned over} waterways, \hl{equipped with} advanced algorithms \hl{designed to detect} and \hl{provide analyses of} potential threats in real-time.
        & The newest unmanned aircraft systems (UAS), \textbf{AI Scan Guards}, \hl{monitor} a physical territory from the air. They \hl{use} image recognition to \hl{analyze} live video streams, \hl{seek out} enemy targets and \hl{alert} the defense forces. 
        & The newest unmanned aircraft systems (UAS), \textbf{AI Scan Guards}, \hl{are programmed to monitor} a physical territory from the air. They \hl{are equipped with} image recognition algorithms that \hl{are used to process} live video streams. \hl{System outputs may be used to identify} enemy targets and \hl{provide alerts to} defense forces.
        \\ \hline
    \end{tabular}}
    \caption{Overview of product descriptions 1-16 (Categories \textit{Properties of a Cognizer} \& \textit{Agency}), Study 1.}
    \label{tab:productdescriptions1}
\end{table}

\begin{table}[]
\small
    \centering
    \renewcommand{\arraystretch}{1.4}%
    \scalebox{0.9}{
    \begin{tabular}{m{0.6cm}|m{4.1cm}|m{4.1cm}|m{4.1cm}|m{4.1cm}}
         \textbf{Cat.} & \textbf{Anthropomorphized} & \textbf{De-anthropomorphized} & \textbf{Anthropomorphized} & \textbf{De-anthropomorpized} \\ \hline

        \rotatebox{90}{\textbf{Biological metaphors}}
        & A software system for court juries, \textbf{Judy}, uses \hl{neural networks} to inform jury members in court cases. Thousands of transcripts and outcomes from previous similar trials are \hl{fed} to Judy's \hl{brain}, whose \hl{digital neurons} \hl{digest} all data and determinants to provide information about relevant law and precedence in current cases. 
        & A software system for court juries, \textbf{Judy}, uses \hl{weighted networks} to inform jury members in court cases. Thousands of transcripts and outcomes from previous similar trials are \hl{input} into Judy's \hl{CPU}, whose \hl{algorithms process} all data and determinants to provide information about relevant law and precedence in current cases.
        
        & A software application, \textbf{JurisDecide} uses \hl{neural networks} to enhance lawyers’ decision-making in trials. Its \hl{digital brain} continually \hl{evolves} and rapidly processes extensive legal data, including precedent and case law which JurisDecide’s \hl{digests} to \hl{spit out} information for legal professionals.  
        & A software application, \textbf{JurisDecide} uses \hl{weighted networks} to enhance lawyers’ decision-making in trials. Its \hl{algorithms} continually \hl{self-update} and rapidly process extensive legal data, including precedent and case law which JurisDecide’s \hl{processes} to \hl{output} information for legal professionals.
        \\ \hline

        \rotatebox{90}{\textbf{Biological metaphors}}
        & A \hl{neural network} system, \textbf{MindHealth}, is an online \hl{digital ear} which \hl{senses} indicators in spoken language that a person may be developing one or more early signs of dementia. Its \hl{digital synapses} have \hl{evolved during} thousands of conversations with healthy humans and dementia patients. & A \hl{weighted network} system, \textbf{MindHealth}, is an online \hl{digital recorder} which \hl{classifies} indicators in spoken language that a person may be developing one or more early signs of dementia. Its \hl{complex algorithms} have \hl{been fine-tuned based on} thousands of conversations with healthy humans and dementia patients.
        
        & A diagnostic tool, \textbf{DermAI Scan}, uses \hl{neural networks} to diagnose dermatological conditions from your home computer. You \hl{feed it} a picture and receive a suggestion for a diagnosis. \hl{Evolving} \hl{neural networks} means that the system’s \hl{neurons} can instantly compare your picture to images of millions of previous diagnoses. 
        & A diagnostic tool, \textbf{DermAI Scan}, uses \hl{weighted networks} to diagnose dermatological conditions from your home computer. You \hl{upload} a picture and receive a suggestion for a diagnosis. \hl{Fine-tuned} \hl{weighted networks} means that the system’s \hl{weights} can instantly compare your picture to images of millions of previous diagnoses.
        \\ \hline

        \rotatebox{90}{\textbf{Properties of a \textit{communicator}}}
        & A smartphone app, \textbf{Lingua}, is an interactive language learning tutor. You can \hl{talk or write to} the app and it will \hl{speak back} to you in real time. Lingua \hl{tells you about} the accuracy and complexity of your speech, and it \hl{suggests} areas of improvement. & A smartphone app, \textbf{Lingua}, is an interactive language learning tutor. You can \hl{input speech or text to} the app and it will \hl{output speech} to you in real time. Lingua \hl{indicates} the accuracy and complexity of your speech, and it \hl{produces suggestions for} areas of improvement.
        
        & \textbf{MentorMe} is an online chatbot, which you can \hl{talk to} about specific academic topics (each based on different data sets). It \hl{speaks like} a mentor, and \hl{proposes} new ways to approach a problem, rather than just \hl{answering} questions directly. It also \hl{asks} you questions to enhance your learning about a given topic. 
        & \textbf{MentorMe} is an online chatbot, into which you can \hl{input text} about specific academic topics (each based on different data sets). It \hl{produces text in the style of} a mentor, and \hl{outputs candidate matches for} new ways to approach a problem, rather than just \hl{indicating answers for} questions directly. It also \hl{outputs} questions to enhance your learning about a given topic.
        \\ \hline

        \rotatebox{90}{\textbf{Properties of a \textit{communicator}}}
        & A smartphone app, \textbf{WardrobEase}, is a service for effortlessly restocking essential clothing items such as jeans, socks, and underwear. It \hl{discusses} your fabric and style preferences \hl{with you}, you \hl{tell it} your sizes, and it \hl{responds with} pictures of choices. You can \hl{tell it} when your clothes are starting to wear out, and \hl{ask it to} \hl{recurringly order} new items from your favorite stores ahead of time. 
        & A smartphone app, \textbf{WardrobEase}, is a service for effortlessly restocking essential clothing items such as jeans, socks, and underwear. It \hl{allows you to record and specify} your fabric and style preferences, you \hl{input} your sizes, and it \hl{outputs} pictures of choices. You can \hl{mark} when your clothes are starting to wear out, and \hl{input automatic, recurring orders} of new items from your favorite stores ahead of time.
        
        & A smartphone app, \textbf{Shoppr}, lets you create meal plans by \hl{discussing} your dietary wishes \hl{with you}. You can \hl{tell the system about} constraints of health, time, nutrition, budget, and it \hl{responds with} suggestions for meals, as well as \hl{write} a meal plan with recipes and \hl{order groceries online for you}. 
        & A smartphone app, \textbf{Shoppr}, lets you create meal plans \hl{based on} your dietary wishes. You can \hl{input} constraints of health, time, nutrition, budget \hl{into the system}, and it \hl{produces} suggestions for meals, as well as \hl{generates} a meal plan with recipes and \hl{an option to put in an online order for groceries}.
        \\ \hline

    \end{tabular}}
    \caption{Overview of product descriptions 17-32 (Categories \textit{Biological metaphors} \& \textit{Properties of a Communicator}), Study 1.}
    \label{tab:productdescriptions2}
\end{table}

\begin{table}[]
\small
    \centering
    \renewcommand{\arraystretch}{1.4}%
    \scalebox{0.9}{
    \begin{tabular}{m{0.6cm}|m{4.1cm}|m{4.1cm}|m{4.1cm}|m{4.1cm}}
         \textbf{Cat.} & \textbf{Anthropomorphized} & \textbf{De-anthropomorphized} & \textbf{Anthropomorphized} & \textbf{De-anthropomorpized} \\ \hline

        \rotatebox{90}{\textbf{Properties of a \textit{cognizer}}}
        & A \hl{machine learning}-based app, \textbf{Lingua}, is an \hl{intelligent} language learning tutor. It \hl{understands} both speech and text and it will produce answers to you in real time. Lingua \hl{identifies} the accuracy and complexity of your speech, and it \hl{recognizes} areas of potential improvement in your spoken language. 
        & A \hl{machine conditioning}-based app, \textbf{Lingua}, is an \hl{automated} language learning tutor. It \hl{processes} both speech and text and it will produce answers to you in real time. Lingua \hl{encodes} the accuracy and complexity of your speech, and it \hl{classifies} areas of potential improvement in your spoken language.
        
        & \textbf{MentorMe} is an \hl{intelligent} online chatbot, with extensive \hl{knowledge} about specific academic topics (each based on different data sets). It \hl{understands} topic-specific questions, and \hl{imagines} new ways to approach a problem, rather than just answering questions directly. It also \hl{comes up with} questions to enhance your learning about a given topic. 
        & \textbf{MentorMe} is an \hl{automated} online chatbot, with extensive \hl{data} about specific academic topics (each based on different data sets). It \hl{processes} topic-specific questions, and \hl{generates text suggesting} new ways to approach a problem, rather than just answering questions directly. It also \hl{produces} questions to enhance your learning about a given topic.
        \\ \hline

        \rotatebox{90}{\textbf{Properties of a \textit{cognizer}}}
        & A phone app, \textbf{WardrobEase}, is an \hl{artificial intelligence}-based service for effortlessly restocking essential clothing items such as jeans, socks, and underwear. It will \hl{learn} your sizes and fabric preferences, and \hl{suggest} pictures of style choices. It \hl{predicts} when clothes are likely to wear out, and can be instructed to \hl{remember} to order new items from your favorite stores ahead of time. 
        & A phone app, \textbf{WardrobEase}, is an \hl{automatic pattern matching} service for effortlessly restocking essential clothing items such as jeans, socks, and underwear. It will \hl{encode} your sizes and fabric preferences, and \hl{display} pictures of style choices. It \hl{makes statistical calculations about} when clothes are likely to wear out, and can be instructed to \hl{automatically} order new items from your favorite stores ahead of time. 
        
        & A \hl{smart} app, \textbf{Shoppr}, lets you create meal plans based on your dietary wishes. It can \hl{remember} constraints about health, time, nutrition, and budget, and \hl{identify} ideas for meals. It can \hl{imagine} monthly meal plans with recipes and \hl{recognize} when to order groceries online. 
        & A \hl{phone} app, \textbf{Shoppr}, lets you create meal plans based on your dietary wishes. It can \hl{encode} constraints about health, time, nutrition, and budget, and \hl{synthesize} ideas for meals. It can \hl{produce} monthly meal plans with recipes and \hl{make statistical predictions about} when to order groceries online.
        \\ \hline

        \rotatebox{90}{\textbf{Agency}}
        & A smartphone app, \textbf{re-Commender}, \hl{collects} data about your dining experiences. It \hl{analyzes} your preferences from historical data, and uses trends from all of its users to \hl{choose} new restaurants you might enjoy. \hl{It stores} your previous ratings and habits, and \hl{picks} offers and coupons you might like. & A smartphone app, \textbf{re-Commender}, \hl{is programmed to collect} data about your dining experiences. \hl{You can use it to analyze} your preferences from historical data, and trends from all of its users to \hl{get suggestions for} new restaurants you might enjoy. \hl{You can save} your previous ratings and habits, and \hl{find} offers and coupons you might like.
        
        & A smartphone app, \textbf{IntelliTrade} is a personal stock broker, which \hl{identifies} promising stocks, funds, and bonds. It \hl{collects} data about your investment preferences as well as historical stock trajectories. It also \hl{analyzes} current news stories, \hl{using} these to \hl{select} promising investment opportunities. 
        & A smartphone app, \textbf{IntelliTrade} is a personal stock broker, which \hl{you can use to identify} promising stocks, funds, and bonds. \hl{It is programmed to store} data about your investment preferences as well as historical stock trajectories. \hl{The algorithms are also frequently run} over current news stories, so \hl{you can use them to find} promising investment opportunities.
        \\ \hline

        \rotatebox{90}{\textbf{Agency}}
        & \textbf{MonAIMaker} is an app that \hl{helps} you plan your personal finances. It \hl{monitors} what you are likely to spend money on by \hl{identifying} trends in your bank statements as well as your email correspondences. It \hl{uses these to find} ways to save money and \hl{remind you} when you have bills and expenses due.
        & \textbf{MonAIMaker} is an app that \hl{you can use to} plan your personal finances. It \hl{is programmed to monitor} what you are likely to spend money on \hl{based on calculations of} trends in your bank statements as well as your email correspondences. \hl{The data can be used to find} ways to save money and \hl{to set up reminders} when you have bills and expenses due.
        
        & An \hl{automatic} app, \textbf{Cameron}, \hl{helps} you organize and answer your emails. It \hl{classifies} text from your incoming emails, and it \hl{suggests} answers based on your writing style. \hl{It also identifies }tasks and deadlines to create automated to-do-lists for you. 
        & An \hl{automatized}  app, \textbf{Cameron}, is a system \hl{you can use to} organize and answer your emails. It \hl{is programmed to classify} text from your incoming emails, and \hl{you can use it to generate} answers based on your writing style. \hl{You can also use it to identify} tasks and deadlines to create automated to-do-lists.
        \\ \hline

    \end{tabular}}
    \caption{Overview of product descriptions 1-16 (Categories \textit{Properties of a Cognizer} \& \textit{Agency}), Study 2.}
    \label{tab:productdescriptions3}
\end{table}

\begin{table}[]
\small
    \centering
    \renewcommand{\arraystretch}{1.4}%
    \scalebox{0.9}{
    \begin{tabular}{m{0.6cm}|m{4.1cm}|m{4.1cm}|m{4.1cm}|m{4.1cm}}
         \textbf{Cat.} & \textbf{Anthropomorphized} & \textbf{De-anthropomorphized} & \textbf{Anthropomorphized} & \textbf{De-anthropomorpized} \\ \hline

        \rotatebox{90}{\textbf{Biological metaphors}}
        & A \hl{neural networks}-based truck \textbf{HaulIT} is made for long haul freight 24/7 without rest stops. The constantly \hl{evolving} algorithms use a \hl{metabolic} principle to minimize their own \hl{synaptic activity} (cost) while maximizing their impact, meaning the truck uses the most efficient routes in both cities and on highways, based on \hl{neural} predictions about traffic conditions and optimal battery charging points. 
        & A \hl{weighted networks}-based truck \textbf{HaulIT} is made for long haul freight 24/7 without rest stops. The constantly \hl{updated} algorithms use an \hl{optimizing} principle to minimize their own \hl{computing activity} (cost) while maximizing their impact, meaning the truck uses the most efficient routes in both cities and on highways, based on \hl{weighted node}-predictions about traffic conditions and optimal battery charging points. 
        
        & A driverless sleeper bus, \textbf{Commuter}, is programmed with \hl{neural networks} to transport people from their home to a long-distance destination overnight. The bus is equipped with \hl{swarm intelligence} to avoid other vehicles and on the road, and the \hl{neural network} adjusts to weather conditions to navigate safely. Its \hl{artificial synapses} are constantly \hl{fed} live traffic data for calculations of the best and safest routes. 
        & A driverless sleeper bus, \textbf{Commuter}, is programmed with \hl{weighted networks} to transport people from their home to a long-distance destination overnight. The bus is equipped with \hl{optimization algorithms} to avoid other vehicles and on the road, and the \hl{weighted network} adjusts to weather conditions to navigate safely. Its \hl{network nodes} are constantly \hl{input} live traffic data for calculations of the best and safest routes.
        \\ \hline

        \rotatebox{90}{\textbf{Biological metaphors}}
        & A \hl{neural network}-powered drone, \textbf{AquaSentinel AI-MAR}, is programmed to passively monitor enemy seas. Equipped with advanced \hl{digital  senses}, it is \hl{watching} and \hl{listening} over waterways. Its \hl{neural network} has been designed to detect and provide analyses of potential threats in real-time. 
        & A \hl{weighted network}-powered drone, \textbf{AquaSentinel AI-MAR}, is programmed to passively monitor enemy seas. Equipped with advanced \hl{digital sensors}, it is \hl{recording video and sound} over waterways. Its \hl{weighted network} has been designed to detect and provide analyses of potential threats in real-time.
        
        & The newest unmanned aircraft systems (UAS), \textbf{AI Scan Guards}, use \hl{neural networks} to passively \hl{watch} a physical territory from the air. Their \hl{neural networks} are specifically trained on image recognition tasks with live video streams, meaning they \hl{see} and \hl{hear} activity instantly. 
        & The newest unmanned aircraft systems (UAS), \textbf{AI Scan Guards}, use \hl{weighted networks} to passively \hl{record video} of a physical territory from the air. Their \hl{weighted networks} are specifically trained on image recognition tasks with live video streams, meaning the predictions \hl{identify image and sound} activity instantly.
        \\ \hline

        \rotatebox{90}{\textbf{Properties of a \textit{communicator}}}
        & A \hl{conversational} software system for court juries, \textbf{Judy}, can be used by jury members to \hl{discuss} active court cases. Judy is based on thousands of transcripts and outcomes from previous similar trials and can \hl{tell} the jury about complex law and precedence. The jury can \hl{ask} Judy to process any kind of data and to \hl{suggest} further avenues of research. All use of Judy is disclosed openly in  court. 
        & A \hl{generative} software system for court juries, \textbf{Judy}, can be used by jury members to \hl{gather information about} active court cases. Judy is based on thousands of transcripts and outcomes from previous similar trials and can \hl{produce text for} the jury about complex law and precedence. The jury can \hl{input} any kind of data into Judy to process and \hl{produce output candidate matches for} further avenues of research. All use of Judy is disclosed openly in  court. 
        
        & A generative AI application, \textbf{JurisDecide}  can be used by lawyers during trials to \hl{speak to} and \hl{debate} their own decision-making processes. JurisDecide is both a source of information and a \hl{chatbot}: it rapidly processes extensive legal data, including precedent and case law, and can both \hl{ask questions of} and \hl{answer questions from} legal professionals.  
        & A generative AI application, \textbf{JurisDecide} can be used by lawyers during trials to \hl{input speech} and \hl{think out loud about} their own decision-making processes. JurisDecide is both a source of information and a \hl{generative text software}: it rapidly processes extensive legal data, including precedent and case law, and can both \hl{produce text in the form of questions and answers} for legal professionals.
        \\ \hline

        \rotatebox{90}{\textbf{Properties of a \textit{communicator}}}
        & \textbf{MindHealth} is an online digital \hl{conversation partner}, which you can \hl{talk to} via your own computer. It classifies indicators in spoken language and can \hl{tell you} if you may be developing one or more early signs of dementia. Its complex algorithms have been fine-tuned based on thousands of conversations with healthy humans and dementia patients, and you can \hl{ask it} questions about its assessment and have it \hl{suggest} further routes of investigation. 
        & \textbf{MindHealth} is an online digital \hl{recorder}, which you can \hl{input speech to} via your own computer. It classifies indicators in spoken language and can \hl{output a statistical prediction} about whether you may be developing one or more early signs of dementia. Its complex algorithms have been fine-tuned based on thousands of conversations with healthy humans and dementia patients, and you can \hl{input} questions about its assessment and have it \hl{output text about} further routes of investigation.
        
        & \textbf{DermAI Scan} uses AI to \hl{respond} to an uploaded picture with \hl{suggestions} for potential dermatological conditions. Fine-tuned algorithms instantly compare your picture to images of millions of previous diagnoses and \hl{tell you} the likelihood of different ones. It can \hl{discuss} different possible diagnoses \hl{with you} if you \hl{tell it} more about the history of your condition. 
        & \textbf{DermAI Scan} uses AI to \hl{generate statistical predictions about} potential dermatological conditions based on an uploaded picture. Fine-tuned algorithms instantly compare your picture to images of millions of previous diagnoses and \hl{output} the likelihood of different ones. It can \hl{generate text about} different possible diagnoses if you \hl{input} more about the history of your condition.
        \\ \hline

    \end{tabular}}
    \caption{Overview of product descriptions 17-32 (Categories \textit{Biological metaphors} \& \textit{Properties of a Communicator}), Study 2.}
    \label{tab:productdescriptions4}
\end{table}
\section{Thematic analysis}
\label{sec:qualitativeanalysis}

A \textbf{thematic analysis} \cite{clarke2015thematic} of the open ended text responses was conducted in the software Condens. All authors went over at least 100 responses and added tags (codes) and notes before a shared discussion about what appeared salient for respondents. All survey responses were read several times while initial codes were generated. The goal of the thematic analysis was to identify patterns that reflect the data for this context \cite{nowell2017thematic}, meaning the goal was to create themes and codes covering all the different responses. The result of the coding was a list of more than 100 different codes at very different levels of abstraction (similarly to the responses, which were also at different level of detail and abstraction). After this, the first author analyzed the remaining responses with codes based on the shared discussions. 

The analysis was an open-ended, inductive treatment, and was focused on \textit{``identifying and interpreting key, but not necessarily all, features of the data, guided by the research question''} \cite{clarke2015thematic}. In practice, each response was read with the overall question in mind: which reason does the respondent provide for being willing to trust or not willing to trust the system? The codes can therefore be seen as `answers' to the research question, such as `accuracy', `reliability', or `risk of bias'. The 30 most prevalent tags are shown in Table \ref{tab:tags}. For an in-depth analysis of the qualitative responses, see \cite{inie2024motivates}.

\begin{table}[h!]
    \centering
    \begin{tabular}{|l|}
         \hline
         perceived usefulness \\
         personal relevance\\
         aversion for other choice\\
         random choice\\
         data quality\\
         expand knowledge/provide guidance\\
         higher accuracy\\
         AI well suited for the task\\
         utilitarianism\\
         volatility of data foundation\\
         no reason\\
         privacy/surveillance\\
         specific product property\\
         impact in case of failure\\
         data type\\
         curiosity/interest/fun/excitement\\
         linguistics\\
         human favoritism\\
         risky, unspecified\\
         larger target market\\
         lower stakes\\
         reliability\\
         augmented humanabilities\\
         individualized/adaptive\\
         AI is unfit for wicked problems\\
         conceptual simplicity\\
         just summarizes description\\
         efficiency \\
         monetary value\\ \hline
    \end{tabular}
    \vspace{1mm}\caption{Tags 
from qualitative responses.}
    \label{tab:tags}
\end{table}

\section{Results}

For all tables, statistically significant $p$-values are indicated in \textbf{bold} and with an $\ast$-symbol. 

\subsection{Results per anthropomorphized category}
\label{sec:appendix_resultscategories}

\begin{table}[h!]
\small 
\renewcommand{\arraystretch}{1.1}
\scalebox{0.95}{
\begin{tabular}{|>{\centering\arraybackslash}m{2.5cm}|>{\centering\arraybackslash}m{0.6cm} >{\centering\arraybackslash}m{0.6cm} >{\centering\arraybackslash}m{1cm} >{\centering\arraybackslash}m{1.4cm} >{\centering\arraybackslash}m{0.6cm} >{\centering\arraybackslash}m{0.6cm}|>{\centering\arraybackslash}m{0.6cm} >{\centering\arraybackslash}m{0.6cm} >{\centering\arraybackslash}m{1cm} >{\centering\arraybackslash}m{1.4cm} >{\centering\arraybackslash}m{0.6cm} >{\centering\arraybackslash}m{0.6cm}|}
\hline
\rowcolor[HTML]{EFEFEF}
\multicolumn{13}{|c|}{\textbf{Study 1 --- Categories of anthropomorphization}} \\ \hline
\textbf{Category} & \multicolumn{6}{c|}{\textbf{Personal trust}} & \multicolumn{6}{c|}{\textbf{General trust}} \\ \hline
 & \textbf{Total} & \textbf{Ant.} & \textbf{De-ant.} & \textbf{\% pref. ant.} & \textbf{\(\chi^2\)} & \textbf{$p$} & \textbf{Total} & \textbf{Ant.} & \textbf{De-ant.} & \textbf{\% pref. ant.} & \textbf{\(\chi^2\)} & \textbf{$p$} \\ \hline

\textbf{Cognizer} & 642 & 363 & 279 & \textbf{56.5} & \textbf{10.99} & \textbf{<.001$\ast$} & 642 & 353 & 289 & \textbf{55.0} & \textbf{6.38} & \textbf{.012$\ast$} \\ \hline

\textbf{Agency} & 628 & 297 & 331 & 47.3 & 1.84 & .17 & 628 & 315 & 313 & 50.2 & 0.01 & 0.94 \\ \hline

\textbf{Bio. metaphors} & 633 & 301 & 332 & 47.6 & 1.61 & .20 & 633 & 301 & 332 & 47.6 & 0.23 & .63 \\ \hline

\textbf{Communication} & 641 & 331 & 310 & 51.6 & 0.624 & .43 & 641 & 326 & 315 & 50.9 & 0.19 & .66 \\ \hline

\multicolumn{13}{|c|}{\textbf{Personal trust: $\chi^2$ = 14.41; $N$ = 2544; $p$ = .002$\ast$}} \\ \hline
\multicolumn{13}{|c|}{General trust: $\chi^2$ 7.26; $N$ = 2493; $p$ = .064} \\ \hline
\end{tabular}}
\caption{Results by categories of anthropomorphization}
\label{tab:categories1}
\end{table}


\begin{table}[h!]
\small 
\renewcommand{\arraystretch}{1.1}
\scalebox{0.95}{
\begin{tabular}{|>{\centering\arraybackslash}m{2.5cm}|>{\centering\arraybackslash}m{0.6cm} >{\centering\arraybackslash}m{0.6cm} >{\centering\arraybackslash}m{1cm} >{\centering\arraybackslash}m{1.4cm} >{\centering\arraybackslash}m{0.6cm} >{\centering\arraybackslash}m{0.6cm}|>{\centering\arraybackslash}m{0.6cm} >{\centering\arraybackslash}m{0.6cm} >{\centering\arraybackslash}m{1cm} >{\centering\arraybackslash}m{1.4cm} >{\centering\arraybackslash}m{0.6cm} >{\centering\arraybackslash}m{0.6cm}|}
\hline
\rowcolor[HTML]{EFEFEF}
\multicolumn{13}{|c|}{\textbf{Study 2 --- Categories of anthropomorphization}} \\ \hline
\textbf{Category} & \multicolumn{6}{c|}{\textbf{Personal trust}} & \multicolumn{6}{c|}{\textbf{General trust}} \\ \hline
 & \textbf{Total} & \textbf{Ant.} & \textbf{De-ant.} & \textbf{\% pref. ant.} & \textbf{\(\chi^2\)} & \textbf{$p$} & \textbf{Total} & \textbf{Ant.} & \textbf{De-ant.} & \textbf{\% pref. ant.} & \textbf{\(\chi^2\)} & \textbf{$p$} \\ \hline

\textbf{Cognizer} & 610 & 304 & 306 & 49.8 & 0.01 & .94 & 607 & 309 & 298 & 50.9 & 0.20 & .66 \\ \hline

\textbf{Agency} & 611 & 336 & 275 & \textbf{55.0} & \textbf{5.89} & \textbf{.015$\ast$} & 609 & 304 & 305 & 49.9 & 0.00 & .97 \\ \hline

\textbf{Bio. metaphors} & 610 & 300 & 310 & 49.2 & 0.16 & .69 & 604 & 307 & 297 & 50.8 & 0.17 & .68 \\ \hline

\textbf{Communication} & 610 & 312 & 298 & 51.1 & 0.32 & .57 & 604 & 314 & 290 & 52.0 & 0.95 & .33 \\ \hline

\multicolumn{13}{|c|}{Personal trust: $\chi^2$ = 4.96; $N$ = 2441; $p$ = .17} \\ \hline
\multicolumn{13}{|c|}{General trust: $\chi^2$ = 0.52; $N$ = 2424; $p$ = .91} \\ \hline
\end{tabular}}
\caption{Results by categories of anthropomorphization}
\label{tab:categories2}
\end{table}


\begin{table}[h!]
\small 
\renewcommand{\arraystretch}{1.1}
\scalebox{0.95}{
\begin{tabular}{|>{\centering\arraybackslash}m{2.5cm}|>{\centering\arraybackslash}m{0.6cm} >{\centering\arraybackslash}m{0.6cm} >{\centering\arraybackslash}m{1cm} >{\centering\arraybackslash}m{1.4cm} >{\centering\arraybackslash}m{0.6cm} >{\centering\arraybackslash}m{0.6cm}|>{\centering\arraybackslash}m{0.6cm} >{\centering\arraybackslash}m{0.6cm} >{\centering\arraybackslash}m{1cm} >{\centering\arraybackslash}m{1.4cm} >{\centering\arraybackslash}m{0.6cm} >{\centering\arraybackslash}m{0.6cm}|}
\hline
\rowcolor[HTML]{EFEFEF}
\multicolumn{13}{|c|}{\textbf{Aggregate results --- Categories of anthropomorphization}} \\ \hline
\textbf{Category} & \multicolumn{6}{c|}{\textbf{Personal trust}} & \multicolumn{6}{c|}{\textbf{General trust}} \\ \hline
 & \textbf{Total} & \textbf{Ant.} & \textbf{De-ant.} & \textbf{\% pref. ant.} & \textbf{\(\chi^2\)} & \textbf{$p$} & \textbf{Total} & \textbf{Ant.} & \textbf{De-ant.} & \textbf{\% pref. ant.} & \textbf{\(\chi^2\)} & \textbf{$p$} \\ \hline

\textbf{Cognizer} & 1252 & 667 & 585 & \textbf{52.3} & \textbf{5.37} & \textbf{.020$\ast$} & 1249 & 662 & 587 & \textbf{53.0} & \textbf{4.50} & \textbf{.034$\ast$} \\ \hline

\textbf{Agency} & 1239 & 633 & 606 & 51.1 & 0.59 & .44 & 1237 & 619 & 618 & 50.0 & 0.00 & .98 \\ \hline

\textbf{Bio. metaphors} & 1243 & 601 & 642 & 48.4 & 1.35 & .24 & 1237 & 608 & 629 & 49.2 & 0.36 & .55 \\ \hline

\textbf{Communication} & 1251 & 643 & 608 & 51.4 & 0.98 & .32 & 1245 & 640 & 605 & 51.4 & 0.98 & .32 \\ \hline

\multicolumn{13}{|c|}{Personal trust: $\chi^2$ = 4.96; $N$ = 4985; $p$ = .17} \\ \hline
\multicolumn{13}{|c|}{General trust: $\chi^2$ = 0.52; $N$ = 4968; $p$ = .91} \\ \hline
\end{tabular}}
\caption{Results by categories of anthropomorphization}
\label{tab:categories3}
\end{table}
\subsection{Results per demographic group, gender}

\begin{table}[h!]
\small 
\renewcommand{\arraystretch}{1.1}
\begin{tabular}{|>{\centering\arraybackslash}m{1.5cm}|>{\centering\arraybackslash}m{0.5cm} >{\centering\arraybackslash}m{0.5cm} >{\centering\arraybackslash}m{1cm} >{\centering\arraybackslash}m{1.4cm} >{\centering\arraybackslash}m{0.5cm} >{\centering\arraybackslash}m{0.5cm}|>{\centering\arraybackslash}m{0.5cm} >{\centering\arraybackslash}m{0.5cm} >{\centering\arraybackslash}m{1cm} >{\centering\arraybackslash}m{1.4cm} >{\centering\arraybackslash}m{0.5cm} >{\centering\arraybackslash}m{0.5cm}|}
\hline
\rowcolor[HTML]{EFEFEF}
\multicolumn{13}{|c|}{\textbf{Study 1 --- Gender}} \\ \hline
\textbf{Gender} & \multicolumn{6}{c|}{\textbf{Personal trust}} & \multicolumn{6}{c|}{\textbf{General trust}} \\ \hline
 & \textbf{Total} & \textbf{Ant.} & \textbf{De-ant.} & \textbf{\% pref. ant.} & \textbf{\(\chi^2\)} & \textbf{$p$} & \textbf{Total} & \textbf{Ant.} & \textbf{De-ant.} & \textbf{\% pref. ant.} & \textbf{\(\chi^2\)} & \textbf{$p$} \\ \hline

\textbf{Female} & 1208 & 592 & 616 & 49.0 & 0.48 & .48 & 1208 & 589 & 616 & 48.8 & 0.60 & .44 \\ \hline

\textbf{Male} & 1264 & 670 & 594 & \textbf{53.0} & \textbf{4.57} & \textbf{.033$\ast$} & 1264 & 672 & 594 & \textbf{53.2} & \textbf{4.81} & \textbf{.028$\ast$} \\ \hline

\textbf{Non-binary} & 33 & 16 & 17 & 48.5 & 0.03 & .86 & 33 & 16 & 17 & 48.5 & 0.03 & .86 \\ \hline

\multicolumn{13}{|c|}{Personal trust: $\chi^2$ = 4.04; $N$ = 2505; $p$ = .13} \\ \hline
\multicolumn{13}{|c|}{General trust: $\chi^2$ = 4.44; $N$ = 2505; $p$ = .11} \\ \hline
\end{tabular}
\caption{Results by self-identified gender}
\label{tab:gender1}
\end{table}


\begin{table}[h!]
\small 
\renewcommand{\arraystretch}{1.1}
\begin{tabular}{|>{\centering\arraybackslash}m{1.5cm}|>{\centering\arraybackslash}m{0.5cm} >{\centering\arraybackslash}m{0.5cm} >{\centering\arraybackslash}m{1cm} >{\centering\arraybackslash}m{1.4cm} >{\centering\arraybackslash}m{0.5cm} >{\centering\arraybackslash}m{0.5cm}|>{\centering\arraybackslash}m{0.5cm} >{\centering\arraybackslash}m{0.5cm} >{\centering\arraybackslash}m{1cm} >{\centering\arraybackslash}m{1.4cm} >{\centering\arraybackslash}m{0.5cm} >{\centering\arraybackslash}m{0.5cm}|}
\hline
\rowcolor[HTML]{EFEFEF}
\multicolumn{13}{|c|}{\textbf{Study 2 --- Gender}} \\ \hline
\textbf{Gender} & \multicolumn{6}{c|}{\textbf{Personal trust}} & \multicolumn{6}{c|}{\textbf{General trust}} \\ \hline
 & \textbf{Total} & \textbf{Ant.} & \textbf{De-ant.} & \textbf{\% pref. ant.} & \textbf{\(\chi^2\)} & \textbf{$p$} & \textbf{Total} & \textbf{Ant.} & \textbf{De-ant.} & \textbf{\% pref. ant.} & \textbf{\(\chi^2\)} & \textbf{$p$} \\ \hline

\textbf{Female} & 1200 & 630 & 570 & 52.5 & 3.00 & .08 & 1200 & 609 & 591 & 50.8 & 0.27 & .60 \\ \hline

\textbf{Male} & 1216 & 621 & 595 & 51.1 & 0.56 & .46 & 1192 & 604 & 588 & 50.7 & 0.21 & .64 \\ \hline

\textbf{Non-binary} & 0 & 0 & 0 & 0 & 0 & 0 & 8 & 4 & 4 & 50.0 & & \\ \hline

\multicolumn{13}{|c|}{Personal trust: $\chi^2$ = 0.49; $N$ = 2416; $p$ = .48} \\ \hline

\multicolumn{13}{|c|}{General trust: $\chi^2$ = .12; $N$ = 2400; $p$ = .94} \\ \hline
\end{tabular}
\caption{Results by self-identified gender}
\label{tab:gender2}
\end{table}


\begin{table}[h!]
\small 
\renewcommand{\arraystretch}{1.1}
\begin{tabular}{|>{\centering\arraybackslash}m{1.5cm}|>{\centering\arraybackslash}m{0.5cm} >{\centering\arraybackslash}m{0.5cm} >{\centering\arraybackslash}m{1cm} >{\centering\arraybackslash}m{1.4cm} >{\centering\arraybackslash}m{0.5cm} >{\centering\arraybackslash}m{0.5cm}|>{\centering\arraybackslash}m{0.5cm} >{\centering\arraybackslash}m{0.5cm} >{\centering\arraybackslash}m{1cm} >{\centering\arraybackslash}m{1.4cm} >{\centering\arraybackslash}m{0.5cm} >{\centering\arraybackslash}m{0.5cm}|}
\hline
\rowcolor[HTML]{EFEFEF}
\multicolumn{13}{|c|}{\textbf{Aggregate results --- Gender}} \\ \hline
\textbf{Gender} & \multicolumn{6}{c|}{\textbf{Personal trust}} & \multicolumn{6}{c|}{\textbf{General trust}} \\ \hline
 & \textbf{Total} & \textbf{Ant.} & \textbf{De-ant.} & \textbf{\% pref. ant.} & \textbf{\(\chi^2\)} & \textbf{$p$} & \textbf{Total} & \textbf{Ant.} & \textbf{De-ant.} & \textbf{\% pref. ant.} & \textbf{\(\chi^2\)} & \textbf{$p$} \\ \hline

\textbf{Female} & 2408 & 1222 & 1186 & 50.7 & 0.54 & .46 & 2408 & 1198 & 1210 & 49.8 & 0.06 & .81 \\ \hline

\textbf{Male} & 2480 & 1291 & 1189 & \textbf{52.1} & \textbf{4.20} & \textbf{.041$\ast$} & 2456 & 1276 & 1180 & 52.0 & 3.75 & .053 \\ \hline

\textbf{Non-binary} & 33 & 16 & 17 & 48.5 & 0.03 & .86 & 41 & 20 & 21 & 48.8 & 0.02 & .88 \\ \hline

\multicolumn{13}{|c|}{Personal trust: $\chi^2$ = 1.00; $N$ = 4921; $p$ = .61} \\ \hline

\multicolumn{13}{|c|}{General trust: $\chi^2$ = 2.43; $N$ = 4905; $p$ = .30} \\ \hline
\end{tabular}
\caption{Results by self-identified gender}
\label{tab:gender3}
\end{table}

\subsection{Results per demographic group, age}

\begin{table}[h!]
\small 
\renewcommand{\arraystretch}{1.1}
\begin{tabular}{|>{\centering\arraybackslash}m{0.8cm}|>{\centering\arraybackslash}m{0.5cm} >{\centering\arraybackslash}m{0.5cm} >{\centering\arraybackslash}m{1cm} >{\centering\arraybackslash}m{1.4cm} >{\centering\arraybackslash}m{0.6cm} >{\centering\arraybackslash}m{0.6cm}|>{\centering\arraybackslash}m{0.5cm} >{\centering\arraybackslash}m{0.5cm} >{\centering\arraybackslash}m{1cm} >{\centering\arraybackslash}m{1.4cm} >{\centering\arraybackslash}m{0.6cm} >{\centering\arraybackslash}m{0.6cm}|}
\hline
\rowcolor[HTML]{EFEFEF}
\multicolumn{13}{|c|}{\textbf{Study 1 --- Age}} \\ \hline
\textbf{Age} & \multicolumn{6}{c|}{\textbf{Personal trust}} & \multicolumn{6}{c|}{\textbf{General trust}} \\ \hline
 & \textbf{Total} & \textbf{Ant.} & \textbf{De-ant.} & \textbf{\% pref. ant.} & \textbf{\(\chi^2\)} & \textbf{$p$} & \textbf{Total} & \textbf{Ant.} & \textbf{De-ant.} & \textbf{\% pref. ant.} & \textbf{\(\chi^2\)} & \textbf{$p$} \\ \hline

\textbf{18-20} & 56 & 27 & 29 & 48.1 & 0.07 & .78 & 56 & 25 & 31 & 44.6 & 0.64 & .42 \\ \hline

\textbf{21-25} & 584 & 290 & 294 & 49.7 & 0.03 & .87 & 584 & 305 & 279 & 52.2 & 1.16 & .28 \\ \hline

\textbf{26-30} & 464 & 224 & 240 & 48.3 & 0.55 & .47 & 464 & 227 & 237 & 48.9 & 0.22 & .64 \\ \hline

\textbf{31-35} & 384 & 212 & 172 & \textbf{55.2} & \textbf{4.17} & \textbf{.043$\ast$} & 384 & 212 & 172 & \textbf{55.2} & \textbf{4.17} & \textbf{.041$\ast$} \\ \hline

\textbf{36-40} & 232 & 119 & 113 & 51.4 & 0.16 & .67 & 232 & 117 & 115 & 50.4 & 0.02 & .90 \\ \hline

\textbf{41-45} & 336 & 169 & 167 & 50.3 & 0.01 & .92 & 336 & 165 & 171 & 49.1 & 0.11 & .74 \\ \hline

\textbf{46-50} & 192 & 94 & 98 & 48.9 & 0.08 & .76 & 192 & 93 & 99 & 48.4 & 0.19 & .66 \\ \hline

\textbf{51-55} & 112 & 57 & 55 & 51.2 & 0.04 & .81 & 112 & 56 & 56 & 50.0 & 0.00 & 1.00 \\ \hline

\textbf{56-60} & 64 & 35 & 30 & 54.8 & 0.38 & .52 & 64 & 28 & 36 & 43.8 & 1.00 & .31 \\ \hline

\textbf{61-65} & 32 & 23 & 9 & \textbf{71.9} & \textbf{7.26} & \textbf{.013$\ast$} & 32 & 21 & 11 & 65.6 & 3.13 & .077 \\ \hline

\textbf{66+} & 56 & 34 & 22 & 60.7 & 2.57 & .11 & 56 & 30 & 26 & 53.6 & 0.29 & .59 \\ \hline

\multicolumn{13}{|c|}{Personal trust: $\chi^2$ = 12.99; $N$ = 2512; $p$ = .22} \\ \hline

\multicolumn{13}{|c|}{General trust: $\chi^2$ = 10.07; $N$ = 2512; $p$ = .43} \\ \hline
\end{tabular}
\caption{Results by age groups.}
\label{tab:age1}
\end{table}


\begin{table}[h!]
\small 
\renewcommand{\arraystretch}{1.1}
\begin{tabular}{|>{\centering\arraybackslash}m{0.8cm}|>{\centering\arraybackslash}m{0.5cm} >{\centering\arraybackslash}m{0.5cm} >{\centering\arraybackslash}m{1cm} >{\centering\arraybackslash}m{1.4cm} >{\centering\arraybackslash}m{0.6cm} >{\centering\arraybackslash}m{0.6cm}|>{\centering\arraybackslash}m{0.5cm} >{\centering\arraybackslash}m{0.5cm} >{\centering\arraybackslash}m{1cm} >{\centering\arraybackslash}m{1.4cm} >{\centering\arraybackslash}m{0.6cm} >{\centering\arraybackslash}m{0.6cm}|}
\hline
\rowcolor[HTML]{EFEFEF}
\multicolumn{13}{|c|}{\textbf{Study 2 --- Age}} \\ \hline
\textbf{Age} & \multicolumn{6}{c|}{\textbf{Personal trust}} & \multicolumn{6}{c|}{\textbf{General trust}} \\ \hline
 & \textbf{Total} & \textbf{Ant} & \textbf{De-ant} & \textbf{\% pref. ant} & \textbf{\(\chi^2\)} & \textbf{$p$} & \textbf{Total} & \textbf{Ant} & \textbf{De-ant} & \textbf{\% pref. ant} & \textbf{\(\chi^2\)} & \textbf{$p$} \\ \hline

\textbf{18-20} & 48 & 20 & 28 & 41.7 & 1.33 & .25 & 24 & 13 & 11 & 54.2 & 0.17 & .68 \\ \hline

\textbf{21-25} & 384 & 201 & 183 & 52.3 & 0.84 & .36 & 336 & 172 & 164 & 51.2 & 0.19 & .66 \\ \hline

\textbf{26-30} & 512 & 274 & 238 & 53.5 & 2.53 & .11 & 416 & 199 & 217 & 47.8 & 0.78 & .38 \\ \hline

\textbf{31-35} & 328 & 163 & 165 & 49.7 & 0.01 & .91 & 416 & 216 & 200 & 51.9 & 0.62 & .43 \\ \hline

\textbf{36-40} & 288 & 161 & 127 & \textbf{55.9} & \textbf{4.01} & \textbf{.045$\ast$} & 248 & 138 & 110 & 55.6 & 3.16 & .075 \\ \hline

\textbf{41-45} & 152 & 71 & 81 & 46.7 & 0.66 & .42 & 104 & 46 & 58 & 44.2 & 1.38 & .24 \\ \hline

\textbf{46-50} & 224 & 121 & 103 & 54.0 & 1.45 & .23 & 272 & 131 & 141 & 48.2 & 0.37 & .54 \\ \hline

\textbf{51-55} & 152 & 74 & 78 & 48.7 & 0.11 & .75 & 232 & 112 & 120 & 48.3 & 0.28 & .60 \\ \hline

\textbf{56-60} & 152 & 73 & 79 & 48.0 & 0.24 & .63 & 128 & 70 & 58 & 54.7 & 1.13 & .29 \\ \hline

\textbf{61-65} & 88 & 58 & 30 & \textbf{65.9} & \textbf{8.91} & \textbf{.003$\ast$} & 88 & 52 & 36 & 59.1 & 2.91 & .08 \\ \hline

\textbf{66+} & 104 & 42 & 62 & 40.4 & 3.85 & .050 & 136 & 68 & 68 & 50.0 & 0.00 & 1.00 \\ \hline

\multicolumn{13}{|c|}{\textbf{Personal trust: $\chi^2$ = 21.06; $N$ = 2432; $p$ = .021$\ast$}} \\ \hline

\multicolumn{13}{|c|}{General trust: $\chi^2$ = 10.49; $N$ = 2400; $p$ = 0.40} \\ \hline
\end{tabular}
\caption{Results by age groups.}
\label{tab:age2}
\end{table}


\begin{table}[h!]
\small 
\renewcommand{\arraystretch}{1.1}
\begin{tabular}{|>{\centering\arraybackslash}m{0.8cm}|>{\centering\arraybackslash}m{0.5cm} >{\centering\arraybackslash}m{0.5cm} >{\centering\arraybackslash}m{1cm} >{\centering\arraybackslash}m{1.4cm} >{\centering\arraybackslash}m{0.6cm} >{\centering\arraybackslash}m{0.6cm}|>{\centering\arraybackslash}m{0.5cm} >{\centering\arraybackslash}m{0.5cm} >{\centering\arraybackslash}m{1cm} >{\centering\arraybackslash}m{1.4cm} >{\centering\arraybackslash}m{0.6cm} >{\centering\arraybackslash}m{0.6cm}|}
\hline
\rowcolor[HTML]{EFEFEF}
\multicolumn{13}{|c|}{\textbf{Aggregate results --- Age}} \\ \hline
\textbf{Age} & \multicolumn{6}{c|}{\textbf{Personal trust}} & \multicolumn{6}{c|}{\textbf{General trust}} \\ \hline
 & \textbf{Total} & \textbf{Ant} & \textbf{De-ant} & \textbf{\% pref. ant} & \textbf{\(\chi^2\)} & \textbf{$p$} & \textbf{Total} & \textbf{Ant} & \textbf{De-ant} & \textbf{\% pref. ant.} & \textbf{\(\chi^2\)} & \textbf{$p$} \\ \hline

\textbf{18-20} & 104 & 47 & 57 & 45.1 & 0.96 & .33 & 80 & 38 & 42 & 47.5 & 0.11 & .74 \\ \hline

\textbf{21-25} & 968 & 491 & 477 & 50.7 & 0.20 & .65 & 920 & 477 & 443 & 51.8 & 1.41 & .23 \\ \hline

\textbf{26-30} & 976 & 498 & 478 & 51.0 & 0.41 & .52 & 880 & 426 & 454 & 48.4 & 0.89 & .34 \\ \hline

\textbf{31-35} & 712 & 375 & 337 & 52.6 & 2.03 & .15 & 800	& 428	& 372	& \textbf{53.5} & \textbf{3.92} & \textbf{.048$\ast$} \\ \hline

\textbf{36-40} & 520 & 280	& 240	& 53.9 & 3.08 & .08 & 480	& 255 & 225	& 53.1 & 1.88 & .17 \\ \hline

\textbf{41-45} & 488 & 240 & 248 & 49.2 & .013 & .71 & 440 & 211 & 229	& 48.0 & 0.74 & .39 \\ \hline

\textbf{46-50} & 416 & 215 & 201 & 51.7 & 0.47 & .49 & 464 & 224 & 240 & 48.3 & 0.55 & .46  \\ \hline

\textbf{51-55} & 264 & 131 & 133 & 49.7 & 0.02 & .90 & 344	& 168 & 176 & 48.8 & 0.19 & .66 \\ \hline

\textbf{56-60} & 216 & 108 & 108 & 50.0 & 0.00 & 1.0 & 192	& 98 & 94 & 51.0 & 0.08 & .77 \\ \hline

\textbf{61-65} & 120 & 81 & 39 & \textbf{67.5} & \textbf{14.70} & \textbf{<.001$\ast$} & 120 & 73 & 47 & 60.8 & \textbf{5.63} & \textbf{.018$\ast$} \\ \hline

\textbf{66+} & 160 & 76 & 84 & 47.5 & 0.40 & .53 & 192 & 98 & 94 & 51.0 & 0.08 & .77 \\ \hline

\multicolumn{13}{|c|}{\textbf{Personal trust: $\chi^2$ 18.45; $N$ = 4944; $p$ = .048$\ast$}} \\ \hline

\multicolumn{13}{|c|}{General trust: $\chi^2$ = 14.51; $N$ = 4912; $p$ = .20} \\ \hline
\end{tabular}
\caption{Results by age groups.}
\label{tab:ageagg}
\end{table}
\subsection{Results per demographic group, socio-economic status}

\begin{table}[h!]
\small 
\scalebox{0.9}{
\renewcommand{\arraystretch}{1.2}
\begin{tabular}{|>{\centering\arraybackslash}m{2.6cm}|>{\centering\arraybackslash}m{0.5cm} >{\centering\arraybackslash}m{0.5cm} >{\centering\arraybackslash}m{1cm} 
>{\centering\arraybackslash}m{1.4cm} >{\centering\arraybackslash}m{0.5cm} >{\centering\arraybackslash}m{0.5cm}|>{\centering\arraybackslash}m{0.5cm} >{\centering\arraybackslash}m{0.5cm} >{\centering\arraybackslash}m{1cm} 
>{\centering\arraybackslash}m{1.4cm} >{\centering\arraybackslash}m{0.5cm} >{\centering\arraybackslash}m{0.5cm}|}
\hline
\rowcolor[HTML]{EFEFEF}
\multicolumn{13}{|c|}{\textbf{Study 1 --- Socio-economic status}} \\ \hline
\textbf{Status} & \multicolumn{6}{c|}{\textbf{Personal trust}} & \multicolumn{6}{c|}{\textbf{General trust}} \\ \hline
 & \textbf{Total} & \textbf{Ant.} & \textbf{De-ant.} & \textbf{\% pref. ant.} & \textbf{\(\chi^2\)} & \textbf{$p$} & \textbf{Total} & \textbf{Ant.} & \textbf{De-ant.} & \textbf{\% pref. ant.} & \textbf{\(\chi^2\)} & \textbf{$p$} \\ \hline

Low & 93 & 48 & 45 & 51.6 & 0.10 & .76 & 93 & 51 & 42 & 54.8 & 0.87 & .35 \\ \hline

Between low/average & 408 & 195 & 213 & 47.8 & 0.79 & .37 & 408 & 199 & 209 & 48.8 & 0.25 & .62 \\ \hline

Average & 1216 & 629 & 587 & 51.7 & 1.45 & .23 & 1216 & 624 & 592 & 51.3 & 0.84 & .36 \\ \hline

Between average/high & 728 & 337 & 391 & \textbf{46.3} & \textbf{4.01} & \textbf{.045$\ast$} & 728 & 379 & 349 & 52.1 & 1.24 & .27 \\ \hline

High & 48 & 23 & 25 & 47.9 & 0.08 & 0.773 & 48 & 20 & 28 & 41.7 & 1.33 & .25 \\ \hline

\multicolumn{13}{|c|}{Personal trust: $\chi^2$ = 6.09; $N$ = 2493; $p$ = .19} \\ \hline
\multicolumn{13}{|c|}{General trust: $\chi^2$ = 3.40; $N$ = 2493; $p$ = .49} \\ \hline
\end{tabular}}
\caption{Results by self-identified socio-economic status.}
\label{tab_socio1}
\end{table}


\begin{table}[h!]
\small 
\scalebox{0.9}{
\renewcommand{\arraystretch}{1.2}
\begin{tabular}{|>{\centering\arraybackslash}m{2.6cm}|>{\centering\arraybackslash}m{0.5cm} >{\centering\arraybackslash}m{0.5cm} >{\centering\arraybackslash}m{1cm} 
>{\centering\arraybackslash}m{1.4cm} >{\centering\arraybackslash}m{0.5cm} >{\centering\arraybackslash}m{0.5cm}|>{\centering\arraybackslash}m{0.5cm} >{\centering\arraybackslash}m{0.5cm} >{\centering\arraybackslash}m{1cm} 
>{\centering\arraybackslash}m{1.4cm} >{\centering\arraybackslash}m{0.5cm} >{\centering\arraybackslash}m{0.5cm}|}
\hline
\rowcolor[HTML]{EFEFEF}
\multicolumn{13}{|c|}{\textbf{Study 2 --- Socio-economic status}} \\ \hline
\textbf{Status} & \multicolumn{6}{c|}{\textbf{Personal trust}} & \multicolumn{6}{c|}{\textbf{General trust}} \\ \hline
 & \textbf{Total} & \textbf{Ant.} & \textbf{De-ant.} & \textbf{\% pref. ant.} & \textbf{\(\chi^2\)} & \textbf{$p$} & \textbf{Total} & \textbf{Ant.} & \textbf{De-ant.} & \textbf{\% pref. ant.} & \textbf{\(\chi^2\)} & \textbf{$p$} \\ \hline

Low & 16 & 11 & 5 & 68.8 & 2.25 & .13 & 32 & 16 & 16 & 50.0 & 0.00 & 1.00 \\ \hline

Between low/average & 256 & 131 & 125 & 51.2 & 0.14 & .71 & 296 & 139 & 157 & 47.0 & 1.09 & .29 \\ \hline

Average & 1264 & 648 & 616 & 51.3 & 0.81 & .37 & 1152 & 612 & 540 & \textbf{53.1} & \textbf{4.50} & \textbf{.034$\ast$} \\ \hline

Between average/high & 848 & 445 & 403 & 52.5 & 2.08 & .15 & 832 & 407 & 425 & 48.9 & 0.39 & .53 \\ \hline

High & 40 & 20 & 20 & 50.0 & 0.00 & 1.00 & 80 & 36 & 44 & 45.0 & 0.80 & .37 \\ \hline

\multicolumn{13}{|c|}{
Personal trust: $\chi^2$ = 2.23; $N$ = 2424; $p$ = .69} \\ \hline
\multicolumn{13}{|c|}{General trust: $\chi^2$ = 6.46; $N$ = 2392; $p$ = .17} \\ \hline
\end{tabular}}
\caption{Results by self-identified socio-economic status.}
\label{tab_socio2}
\end{table}


\begin{table}[h!]
\small 
\scalebox{0.9}{
\renewcommand{\arraystretch}{1.2}
\begin{tabular}{|
>{\centering\arraybackslash}m{2.6cm}|>{\centering\arraybackslash}m{0.5cm} >{\centering\arraybackslash}m{0.5cm} >{\centering\arraybackslash}m{1cm} 
>{\centering\arraybackslash}m{1.4cm} >{\centering\arraybackslash}m{0.5cm} >{\centering\arraybackslash}m{0.5cm}|>{\centering\arraybackslash}m{0.5cm} >{\centering\arraybackslash}m{0.5cm} >{\centering\arraybackslash}m{1cm} 
>{\centering\arraybackslash}m{1.4cm} >{\centering\arraybackslash}m{0.5cm} >{\centering\arraybackslash}m{0.5cm}|}
\hline
\rowcolor[HTML]{EFEFEF}
\multicolumn{13}{|c|}{\textbf{Aggregate results --- Socio-economic status}} \\ \hline
\textbf{Status} & \multicolumn{6}{c|}{\textbf{Personal trust}} & \multicolumn{6}{c|}{\textbf{General trust}} \\ \hline
 & \textbf{Total} & \textbf{Ant.} & \textbf{De-ant.} & \textbf{\% pref. ant.} & \textbf{\(\chi^2\)} & \textbf{$p$} & \textbf{Total} & \textbf{Ant.} & \textbf{De-ant.} & \textbf{\% pref. ant.} & \textbf{\(\chi^2\)} & \textbf{$p$} \\ \hline

Low & 109 & 59 & 50 & 54.1 & 0.74  & .39 & 
125 & 67 & 58 & 53.6 & 0.65  &  .42 \\ \hline

Between low/average & 664 & 326 & 338 & 49.1 & 0.22  & .64  & 
704 & 338 & 366 & 48.0 &  1.11 & .29  \\ \hline

Average & 2480 & 1277 & 1203 & 51.5 & 2.21  & .14 & 
2368 & 1236 & 1132 & \textbf{52.2} & \textbf{4.57}  & \textbf{.034$\ast$} \\ \hline

Between average/high & 1576 & 782 & 794 & 49.6 & 0.09 & .76  &   
1560 & 786 & 774 & 50.4  & 0.09  & .76  \\ \hline

High & 88 & 43 & 45 & 48.9 & 0.05 & .83 &
128 & 56 & 72 & 43.8  & 2.00  & .16 \\ \hline

\multicolumn{13}{|c|}{Personal trust: $\chi^2$ = 2.64; $N$ = 4917; $p$ = .62} \\ \hline
\multicolumn{13}{|c|}{General trust: $\chi^2$ = 7.08; $N$ = 4885; $p$ = .13} \\ \hline
\end{tabular}}
\caption{Results by self-identified socio-economic status.}
\label{tab_socioagg}
\end{table}
\subsection{Results per demographic group, education}

\begin{table}[h!]
\small 
\setlength\tabcolsep{3pt} 
\begin{tabular}{|>{\centering\arraybackslash}m{2.5cm}|>{\centering\arraybackslash}m{0.6cm} >{\centering\arraybackslash}m{0.6cm} >{\centering\arraybackslash}m{1cm} 
>{\centering\arraybackslash}m{1.4cm} >{\centering\arraybackslash}m{0.6cm} >{\centering\arraybackslash}m{0.6cm}|>{\centering\arraybackslash}m{0.6cm} >{\centering\arraybackslash}m{0.6cm} >{\centering\arraybackslash}m{1cm} 
>{\centering\arraybackslash}m{1.4cm} >{\centering\arraybackslash}m{0.6cm} >{\centering\arraybackslash}m{0.6cm}|}
\hline
\rowcolor[HTML]{EFEFEF}
\multicolumn{13}{|c|}{\textbf{Study 1 --- Educational level}} \\ \hline
\textbf{Level} & \multicolumn{6}{c|}{\textbf{Personal trust}} & \multicolumn{6}{c|}{\textbf{General trust}} \\ \hline
 & \textbf{Total} & \textbf{Ant.} & \textbf{De-ant.} & \textbf{\% pref. ant.} & \textbf{\(\chi^2\)} & \textbf{$p$} & \textbf{Total} & \textbf{Ant.} & \textbf{De-ant.} & \textbf{\% pref. ant.} & \textbf{\(\chi^2\)} & \textbf{$p$} \\ \hline
 
 \textbf{High school}  & 584 & 285 & 299 & 48.8 & 0.34 & .56 & 584 & 286 & 298 & 49.0 & 0.25 & .62 \\ \hline
 
 \textbf{Bachelor's}  & 1120 & 564 & 556 & 50.4 & 0.06 & .81 & 1120 & 573 & 547 & 51.2 & 0.60 & .44 \\ \hline
 
 \textbf{Master's}  & 616 & 337 & 279 & \textbf{54.7} & \textbf{5.46} & \textbf{.019$\ast$} & 616 & 324 & 292 & 52.6 & 1.66 & .20 \\ \hline
 
 \textbf{PhD}  & 144 & 74 & 70 & 51.4 & 0.11 & .74 & 144 & 77 & 67 & 53.5 & 0.69 & .40 \\ \hline

 \textbf{None/No answer}  & 49 & 19 & 30 & 38.8 & 2.47 & .12 & 49 & 19 & 30 & 38.8 & 2.47 & .12 \\ \hline

\multicolumn{13}{|c|}{Personal trust: $\chi^2$ 7.63; $N$ = 2432; $p$ = .11} \\ \hline

\multicolumn{13}{|c|}{General trust: $\chi^2$ = 4.87; $N$ = 2400; $p$ = .30} \\ \hline
\end{tabular}
\caption{Results by educational level.}
\label{tab:education1}
\end{table}


\begin{table}[h!]
\small 
\setlength\tabcolsep{3pt} 
\begin{tabular}{|>{\centering\arraybackslash}m{2.5cm}|>{\centering\arraybackslash}m{0.6cm} >{\centering\arraybackslash}m{0.6cm} >{\centering\arraybackslash}m{1cm} 
>{\centering\arraybackslash}m{1.4cm} >{\centering\arraybackslash}m{0.6cm} >{\centering\arraybackslash}m{0.6cm}|>{\centering\arraybackslash}m{0.6cm} >{\centering\arraybackslash}m{0.6cm} >{\centering\arraybackslash}m{1cm} 
>{\centering\arraybackslash}m{1.4cm} >{\centering\arraybackslash}m{0.6cm} >{\centering\arraybackslash}m{0.6cm}|}
\hline
\rowcolor[HTML]{EFEFEF}
\multicolumn{13}{|c|}{\textbf{Study 2 --- Educational level}} \\ \hline
\textbf{Level} & \multicolumn{6}{c|}{\textbf{Personal trust}} & \multicolumn{6}{c|}{\textbf{General trust}} \\ \hline
 & \textbf{Total} & \textbf{Ant.} & \textbf{De-ant.} & \textbf{\% pref. ant.} & \textbf{\(\chi^2\)} & \textbf{$p$} & \textbf{Total} & \textbf{Ant.} & \textbf{De-ant.} & \textbf{\% pref. ant.} & \textbf{\(\chi^2\)} & \textbf{$p$} \\ \hline

\textbf{High school}  & 504 & 247 & 257 & 49.0 & 0.20 & .66 & 504 & 252 & 252 & 50.0 & 0.00 & 1.00 \\ \hline

\textbf{Bachelor's} & 1104 & 579 & 525 & 52.4 & 2.64 & .10 & 960 & 503 & 457 & 52.4 & 2.20 & .14 \\ \hline

\textbf{Master's}  Degree & 744 & 398 & 346 & 53.5 & 3.63 & .057 & 456 & 227 & 229 & 49.8 & 0.01 & .92 \\ \hline

\textbf{PhD}  & 72 & 32 & 40 & 44.4 & 0.89 & .35 & 432 & 211 & 221 & 48.8 & 0.23 & .63 \\ \hline

\textbf{None/No answer} & 8 & 2 & 6 & 25.0 & & & 48 & 24 & 24 & 50.0 & 0.00 & 1.00 \\ \hline

\multicolumn{13}{|c|}{Personal trust: $\chi^2$ = 4.17; $N$ = 2432; $p$ = .24} \\ \hline

\multicolumn{13}{|c|}{General trust: $\chi^2$ = 1.96; $N$ = 2400; $p$ = .74} \\ \hline
\end{tabular}
\caption{Results by educational level.}
\label{tab:education2}
\end{table}


\begin{table}[h!]
\small 
\setlength\tabcolsep{3pt} 
\begin{tabular}{|>{\centering\arraybackslash}m{2.5cm}|>{\centering\arraybackslash}m{0.6cm} >{\centering\arraybackslash}m{0.6cm} >{\centering\arraybackslash}m{1cm} 
>{\centering\arraybackslash}m{1.4cm} >{\centering\arraybackslash}m{0.6cm} >{\centering\arraybackslash}m{0.6cm}|>{\centering\arraybackslash}m{0.6cm} >{\centering\arraybackslash}m{0.6cm} >{\centering\arraybackslash}m{1cm} 
>{\centering\arraybackslash}m{1.4cm} >{\centering\arraybackslash}m{0.6cm} >{\centering\arraybackslash}m{0.6cm}|}
\hline
\rowcolor[HTML]{EFEFEF}
\multicolumn{13}{|c|}{\textbf{Aggregate results --- Educational level}} \\ \hline
\textbf{Level} & \multicolumn{6}{c|}{\textbf{Personal trust}} & \multicolumn{6}{c|}{\textbf{General trust}} \\ \hline
 & \textbf{Total} & \textbf{Ant.} & \textbf{De-ant.} & \textbf{\% pref. ant.} & \textbf{\(\chi^2\)} & \textbf{$p$} & \textbf{Total} & \textbf{Ant.} & \textbf{De-ant.} & \textbf{\% pref. ant.} & \textbf{\(\chi^2\)} & \textbf{$p$} \\ \hline

 \textbf{High school}  & 1088 & 532 & 556 & 48.9 & 0.53 & .47 & 
1088 & 538 & 550 & 49.4 & 0.13 & .72   \\ \hline

\textbf{Bachelor's} & 2224 & 1173 & 1081 & 51.4 & 3.76 & .053 & 
2080 & 1076 & 1007 & 51.7 & 2.29 & .13 \\ \hline

 \textbf{Master's} & 1360 & 735 & 625 & \textbf{54.0} & \textbf{8.90} & \textbf{.003$\ast$} & 
1072 & 551 & 521 & 51.4 & .084 & .36 \\ \hline

 \textbf{PhD}  & 216 & 106 & 110 & 49.1 & 0.07 & .78 &
576 & 288 & 288 & 50.0 & 0.00 & 1.00 \\ \hline

\textbf{None/No answer} & 8 & 2 & 6 & 25.0 & N/A\textsuperscript{1} & .29 & 48 & 24 & 24 & 50.0 & 0.00 & 1.00 \\ \hline

\multicolumn{13}{|c|}{Personal trust: $\chi^2$ = 6.99; $N$ = 4888; $p$ = .07, not significant at \( p < .05 \)} \\ \hline

\multicolumn{13}{|c|}{General trust: $\chi^2$ = 1.78; $N$ = 4816; $p$ = .62, not significant at \( p < .05 \)} \\ \hline
\end{tabular}
\caption{Results by educational level. \textsuperscript{1}Because of the low $N$, a Fisher Exact test was performed on these numbers.}
\label{tab:educationagg}
\end{table}

\subsection{Results per demographic group, self-estimated computer knowledge}


\begin{table}[h!]
\small 
\renewcommand{\arraystretch}{1.1}
\setlength\tabcolsep{3pt} 
\begin{tabular}{
|>{\centering\arraybackslash}m{2.5cm}|
>{\centering\arraybackslash}m{0.6cm} >{\centering\arraybackslash}m{0.6cm} >{\centering\arraybackslash}m{1cm} >{\centering\arraybackslash}m{1.4cm} >{\centering\arraybackslash}m{0.6cm} >{\centering\arraybackslash}m{0.6cm}|>{\centering\arraybackslash}m{0.6cm} >{\centering\arraybackslash}m{0.6cm} >{\centering\arraybackslash}m{1cm} >{\centering\arraybackslash}m{1.4cm} >{\centering\arraybackslash}m{0.6cm} >{\centering\arraybackslash}m{0.6cm}|}
\hline
\rowcolor[HTML]{EFEFEF}
\multicolumn{13}{|c|}{\textbf{Study 1 --- Computer knowledge}} \\ \hline
\textbf{Knowledge} & \multicolumn{6}{c|}{\textbf{Personal trust}} & \multicolumn{6}{c|}{\textbf{General trust}} \\ \hline
 & \textbf{Total} & \textbf{Ant.} & \textbf{De-ant.} & \textbf{\% pref. ant.} & \textbf{\(\chi^2\)} & \textbf{$p$} & \textbf{Total} & \textbf{Ant.} & \textbf{De-ant.} & \textbf{\% pref. ant.} & \textbf{\(\chi^2\)} & \textbf{$p$} \\ \hline

Lower than av. & 40 & 22 & 18 & 55.0 & 0.40 & .53 & 40 & 21 & 19 & 52.5 & 0.10 & .75 \\ \hline

Average & 952 & 486 & 466 & 51.1 & 0.42 & .52 & 952 & 483 & 469 & 50.7 & 0.21 & .65 \\ \hline

Higher than av. & 1256 & 646 & 610 & 51.4 & 1.03 & .31 & 1256 & 646 & 610 & 51.4 & 1.03 & .31 \\ \hline

High (can code) & 256 & 122 & 134 & 47.7 & 0.56 & .45 & 256 & 127 & 129 & 49.6 & 0.02 & .90 \\ \hline

\multicolumn{13}{|c|}{Personal trust: $\chi^2$ = 1.4949; $N$ = 2504; $p$ = .68} \\ \hline

\multicolumn{13}{|c|}{General trust: $\chi^2$ = 0.35; $N$ = 2504; $p$ = .95} \\ \hline

\end{tabular}
\caption{Results by self-estimated level of computer knowledge.}
\label{tab:comp1}
\end{table}


\begin{table}[h!]
\small 
\renewcommand{\arraystretch}{1.1}
\setlength\tabcolsep{3pt} 
\begin{tabular}{
|>{\centering\arraybackslash}m{2.5cm}|
>{\centering\arraybackslash}m{0.6cm} >{\centering\arraybackslash}m{0.6cm} >{\centering\arraybackslash}m{1cm} >{\centering\arraybackslash}m{1.4cm} >{\centering\arraybackslash}m{0.6cm} >{\centering\arraybackslash}m{0.6cm}|>{\centering\arraybackslash}m{0.6cm} >{\centering\arraybackslash}m{0.6cm} >{\centering\arraybackslash}m{1cm} >{\centering\arraybackslash}m{1.4cm} >{\centering\arraybackslash}m{0.6cm} >{\centering\arraybackslash}m{0.6cm}|}
\hline
\rowcolor[HTML]{EFEFEF}
\multicolumn{13}{|c|}{\textbf{Study 2 --- Computer knowledge}} \\ \hline
\textbf{Knowledge Level} & \multicolumn{6}{c|}{\textbf{Personal trust}} & \multicolumn{6}{c|}{\textbf{General trust}} \\ \hline
 & \textbf{Total} & \textbf{Ant.} & \textbf{De-ant.} & \textbf{\% pref. ant.} & \textbf{\(\chi^2\)} & \textbf{$p$} & \textbf{Total} & \textbf{Ant.} & \textbf{De-ant.} & \textbf{\% pref. ant.} & \textbf{\(\chi^2\)} & \textbf{$p$} \\ \hline

Lower than av. & 64 & 30 & 34 & 46.9 & 0.25 & .62 & 80 & 40 & 40 & 50.0 & 0.00 & 1.00 \\ \hline

Average & 952 & 502 & 450 & 52.7 & 2.84 & .09 & 792 & 396 & 396 & 50.0 & 0.00 & 1.00 \\ \hline

Higher than av. & 1144 & 591 & 553 & 51.7 & 1.26 & .26 & 1232 & 632 & 600 & 51.3 & 0.83 & .36 \\ \hline

High (can code) & 264 & 129 & 135 & 48.9 & 0.14 & .71 & 288 & 143 & 145 & 49.7 & 0.01 & .91 \\ \hline

\multicolumn{13}{|c|}{Personal trust: $\chi^2$ = 1.85; $N$ = 2432; $p$ = .60} \\ \hline
\multicolumn{13}{|c|}{General trust: $\chi^2$ = 0.47; $N$ = 2400; $p$ = .93} \\ \hline
\end{tabular}
\caption{Results by self-estimated level of computer knowledge.}
\label{tab:comp2}
\end{table}


\begin{table}[h!]
\small 
\renewcommand{\arraystretch}{1.1}
\setlength\tabcolsep{3pt} 
\begin{tabular}{
|>{\centering\arraybackslash}m{2.5cm}|
>{\centering\arraybackslash}m{0.6cm} >{\centering\arraybackslash}m{0.6cm} >{\centering\arraybackslash}m{1cm} >{\centering\arraybackslash}m{1.4cm} >{\centering\arraybackslash}m{0.6cm} >{\centering\arraybackslash}m{0.6cm}|>{\centering\arraybackslash}m{0.6cm} >{\centering\arraybackslash}m{0.6cm} >{\centering\arraybackslash}m{1cm} >{\centering\arraybackslash}m{1.4cm} >{\centering\arraybackslash}m{0.6cm} >{\centering\arraybackslash}m{0.6cm}|}
\hline
\rowcolor[HTML]{EFEFEF}
\multicolumn{13}{|c|}{\textbf{Aggregate results --- Computer knowledge}} \\ \hline
\textbf{Knowledge Level} & \multicolumn{6}{c|}{\textbf{Personal trust}} & \multicolumn{6}{c|}{\textbf{General trust}} \\ \hline
 & \textbf{Total} & \textbf{Ant.} & \textbf{De-ant.} & \textbf{\% pref. ant.} & \textbf{\(\chi^2\)} & \textbf{$p$} & \textbf{Total} & \textbf{Ant.} & \textbf{De-ant.} & \textbf{\% pref. ant.} & \textbf{\(\chi^2\)} & \textbf{$p$} \\ \hline

Lower than av. & 104 & 52 & 52 & 50.0 & 0.00 & 1.00 &
120 & 61 & 59 & 50.8 & 0.03 & .85 \\ \hline

Average & 1904 & 988 & 916 & 51.9 & 2.72 & .09 &
1744 & 879 & 865 & 50.4 & 0.11 & .74 \\ \hline

Higher than av. & 2400 & 1237 & 1163 & 51.5 & 2.28 & .13 &
2488 & 1278 & 1210 & 51.4 & 1.86 & .17 \\ \hline

High (can code) & 520 & 250 & 269 & 48.3 & 0.70 & .40 &
544 & 270 & 274 & 49.6 & 0.03 & .86 \\ \hline

\multicolumn{13}{|c|}{Personal trust: $\chi^2$ = 2.30; $N$ = 4928; $p$ = .51} \\ \hline
\multicolumn{13}{|c|}{General trust: $\chi^2$ = 4.72; $N$ = 4896; $p$ = .32} \\ \hline
\end{tabular}
\caption{Results by self-estimated level of computer knowledge.}
\label{tab:comp3}
\end{table}
\section{Demographics}
\label{sec:demographics}

\subsection{Study 1}

\begin{figure}[h!]
    \centering
    \includegraphics[width=7cm]{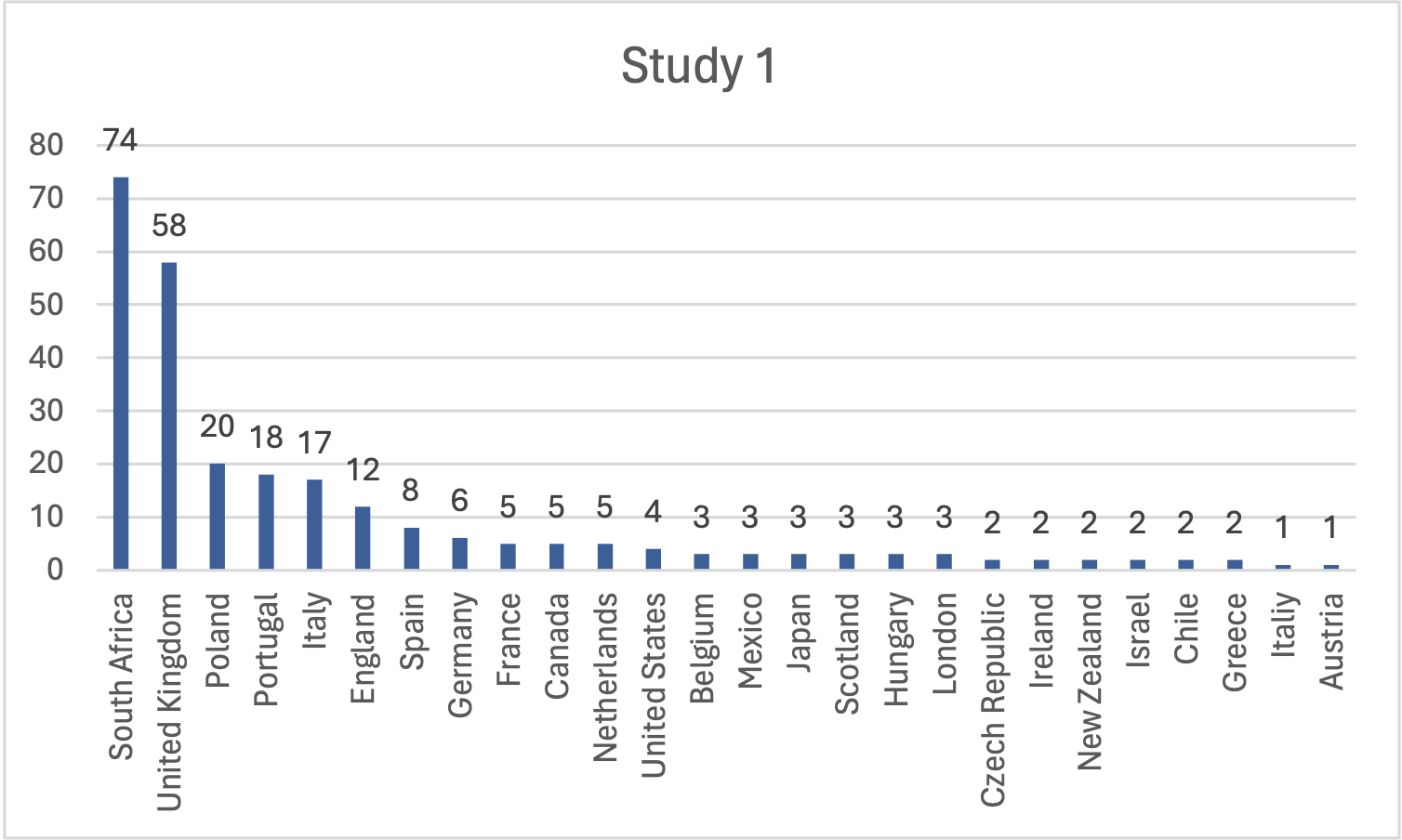}
    \caption{Overview of countries of residence of participants. Note, that these numbers do not equal the full amount of participants, since this was an open-text-question, and not all participants provided a (useful) answer.}
    \label{fig:countries1}
\end{figure}

\begin{table}[h!]
\small 
\centering
\scalebox{0.9}{
\begin{tabular}{cc}
\begin{tabular}{|l|l|l|}
\hline
\textbf{Age Group} & \textbf{Percent} & \textbf{\#} \\ \hline
18-20 & 2.2\% & 7 \\ \hline
21-25 & 23.2\% & 73 \\ \hline
26-30 & 18.5\% & 58 \\ \hline
31-35 & 15.3\% & 48 \\ \hline
36-40 & 9.2\% & 29 \\ \hline
41-45 & 13.4\% & 42 \\ \hline
46-50 & 7.6\% & 24 \\ \hline
51-55 & 4.5\% & 14 \\ \hline
56-60 & 2.5\% & 8 \\ \hline
61-65 & 1.3\% & 4 \\ \hline
66+ & 2.2\% & 7 \\ \hline
\textbf{Total} & 100\% & 314 \\ \hline
\end{tabular}
\quad
\begin{tabular}{|l|l|l|}
\hline
\textbf{Gender} & \textbf{Percent} & \textbf{\#} \\ \hline
Female & 48.1\% & 151 \\ \hline
Non-binary & 1.3\% & 4 \\ \hline
Male & 50.3\% & 158 \\ \hline
Prefer not to answer & 0.3\% & 1 \\ \hline
\textbf{Total} & 100\% & 314 \\ \hline
\end{tabular}
\quad
\begin{tabular}{|l|l|l|}
\hline
\textbf{Race or Ethnicity} & \textbf{Percent} & \textbf{\#} \\ \hline
American Indian or Alaskan Native & 0\% & 0 \\ \hline
Asian / Pacific Islander & 11.1\% & 35 \\ \hline
Black or African American & 29\% & 91 \\ \hline
Hispanic / Latina/o & 3.8\% & 12 \\ \hline
White / Caucasian & 45.5\% & 143 \\ \hline
Multiple ethnicity / Other & 8.6\% & 27 \\ \hline
Prefer not to answer & 1.9\% & 6 \\ \hline
\textbf{Total} & 100\% & 314 \\ \hline
\end{tabular}
\\
\begin{tabular}{|l|l|l|}
\hline
\textbf{Socio-economic Status} & \textbf{Percent} & \textbf{\#} \\ \hline
Low & 3.8\% & 12 \\ \hline
Between Low and Average & 16.3\% & 51 \\ \hline
Average & 48.7\% & 152 \\ \hline
Between Average and High & 29.2\% & 91 \\ \hline
High & 1.9\% & 6 \\ \hline
\textbf{Total} & 100\% & 312 \\ \hline
\end{tabular}
\end{tabular}}

\caption{Demographics: Age, gender, race or ethnicity, and socio-economic status}
\label{demo_age_gend_eth_soc1}
\end{table}

\begin{table}[h!]
\small
\centering
\scalebox{0.9}{
\begin{tabular}{cc}
\begin{tabular}{|l|l|l|}
\hline
\textbf{Education Level} & \textbf{Percent} & \textbf{\#} \\ \hline
High School or Equivalent & 23.2\% & 73 \\ \hline
Bachelors Degree or Equivalent & 44.6\% & 140 \\ \hline
Masters Degree or Equivalent & 24.5\% & 77 \\ \hline
PhD or Equivalent & 5.7\% & 18 \\ \hline
None / Prefer not to answer & 1.9\% & 6 \\ \hline
\textbf{Total} & 100\% & 314 \\ \hline
\end{tabular}
&
\begin{tabular}{|l|l|l|}
\hline
\textbf{Computer Knowledge} & \textbf{Percent} & \textbf{\#} \\ \hline
Low (Rarely use computers) & 0\% & 0 \\ \hline
Lower than Average & 1.6\% & 5 \\ \hline
Average & 38.0\% & 119 \\ \hline
Higher than Average & 50.1\% & 157 \\ \hline
High (Can code) & 10.2\% & 32 \\ \hline
\textbf{Total} & 100\% & 313 \\ \hline
\end{tabular}
\end{tabular}}

\caption{Demographics: Education level and computer knowledge}
\label{demo_edu_comp1}
\label{demo_education1}
\end{table}

\clearpage

\subsection{Study 2}

\begin{figure}[h!]
    \centering
    \includegraphics[width=9cm]{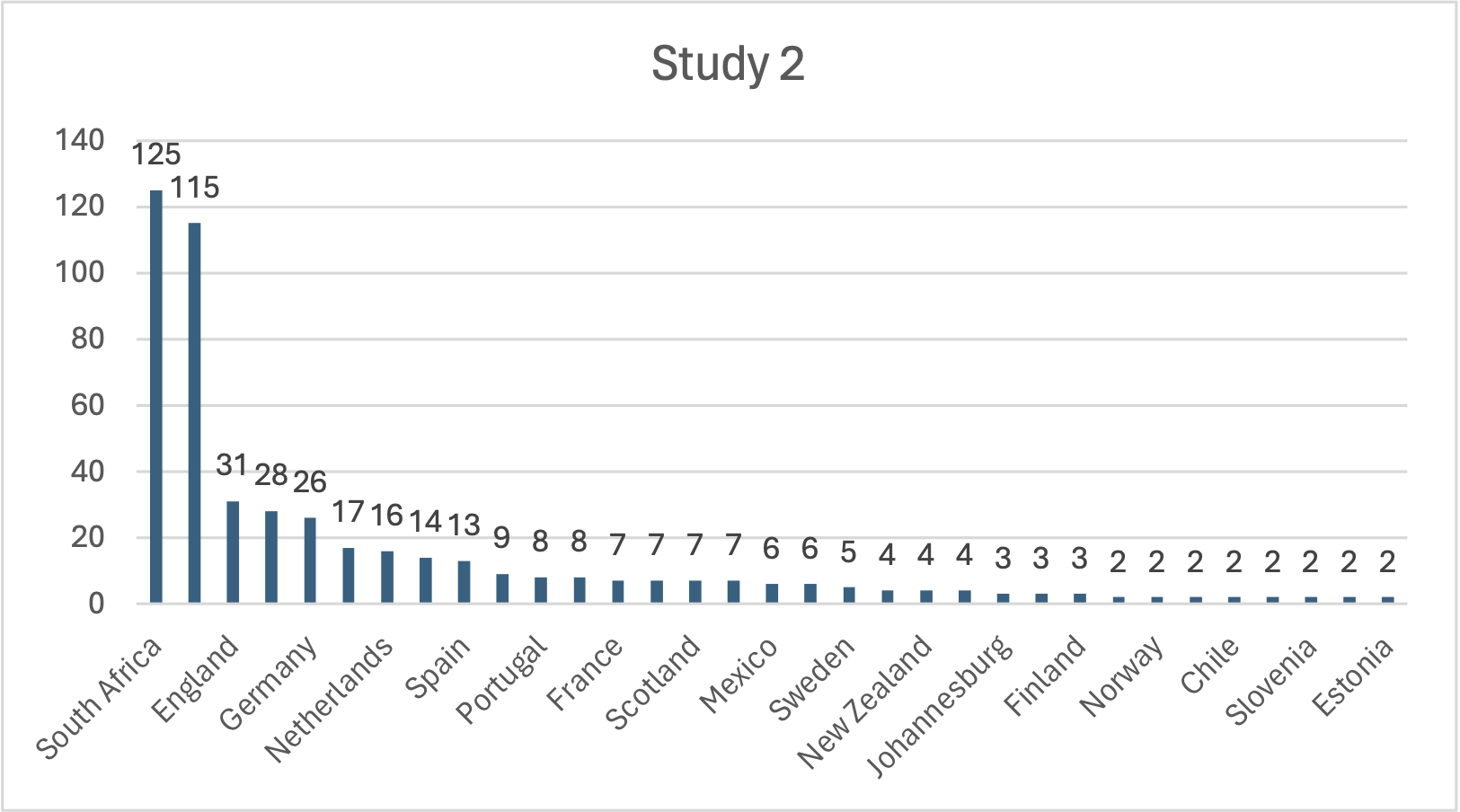}
    \caption{Overview of countries of residence of participants. Note, that these numbers do not equal the full amount of participants, since this was an open-text-question, and not all participants provided a (useful) answer.}
    \label{fig:countries2}
\end{figure}

\begin{table}[h!]
\small 
\centering
\scalebox{0.9}{
\begin{tabular}{cc}
\begin{tabular}{|l|l|l|}
\hline
\textbf{Age Group} & \textbf{Percent} & \textbf{\#} \\ \hline
18-20 & 1.5\% & 9 \\ \hline
21-25 & 14.9\% & 90 \\ \hline
26-30 & 19.2\% & 116 \\ \hline
31-35 & 15.4\% & 93 \\ \hline
36-40 & 11.1\% & 67 \\ \hline
41-45 & 5.3\% & 32 \\ \hline
46-50 & 10.3\% & 62 \\ \hline
51-55 & 7.9\% & 48 \\ \hline
56-60 & 5.8\% & 35 \\ \hline
61-65 & 3.6\% & 22 \\ \hline
66+ & 5\% & 30 \\ \hline
\textbf{Total} & 100\% & 604 \\ \hline
\end{tabular}
\quad
\begin{tabular}{|l|l|l|}
\hline
\textbf{Gender} & \textbf{Percent} & \textbf{\#} \\ \hline
Female & 49.7\% & 300 \\ \hline
Non-binary & 0.2\% & 1 \\ \hline
Male & 49.8\% & 301 \\ \hline
Prefer not to answer & 0.3\% & 2 \\ \hline
\textbf{Total} & 100\% & 604 \\ \hline
\end{tabular}
\quad
\begin{tabular}{|l|l|l|}
\hline
\textbf{Race or Ethnicity} & \textbf{Percent} & \textbf{\#} \\ \hline
American Indian or Alaskan Native & 0.2\% & 1 \\ \hline
Asian / Pacific Islander & 6.8\% & 41 \\ \hline
Black or African American & 27.2\% & 164 \\ \hline
Hispanic / Latina/o & 4.1\% & 25 \\ \hline
White / Caucasian & 55\% & 332 \\ \hline
Multiple ethnicity / Other & 5.6\% & 34 \\ \hline
Prefer not to answer & 1.2\% & 7 \\ \hline
\textbf{Total} & 100\% & 314 \\ \hline
\end{tabular}
\\
\begin{tabular}{|l|l|l|}
\hline
\textbf{Socio-economic Status} & \textbf{Percent} & \textbf{\#} \\ \hline
Low & 1\% & 6 \\ \hline
Between Low and Average & 11.5\% & 69 \\ \hline
Average & 50.2\% & 302 \\ \hline
Between Average and High & 34.9\% & 210 \\ \hline
High & 2.5\% & 15 \\ \hline
\textbf{Total} & 100\% & 602 \\ \hline
\end{tabular}
\end{tabular}}

\caption{Demographics: Age, gender, race or ethnicity, and socio-economic status}
\label{demo_age_gend_eth_soc2}
\end{table}

\begin{table}[h!]
\small
\centering
\scalebox{0.9}{
\begin{tabular}{cc}
\begin{tabular}{|l|l|l|}
\hline
\textbf{Education Level} & \textbf{Percent} & \textbf{\#} \\ \hline
High School or Equivalent & 20.9\% & 126 \\ \hline
Bachelors Degree or Equivalent & 42.7\% & 258 \\ \hline
Masters Degree or Equivalent & 24.8\% & 150 \\ \hline
PhD or Equivalent & 10.4\% & 63 \\ \hline
None / Prefer not to answer & 1.2\% & 7 \\ \hline
\textbf{Total} & 100\% & 314 \\ \hline
\end{tabular}
&
\begin{tabular}{|l|l|l|}
\hline
\textbf{Computer Knowledge} & \textbf{Percent} & \textbf{\#} \\ \hline
Lower than Average & 3.3\% & 20 \\ \hline
Average & 36.1\% & 218 \\ \hline
Higher than Average & 49.2\% & 297 \\ \hline
High (Can code) & 11.4\% & 69 \\ \hline
\textbf{Total} & 100\% & 604 \\ \hline
\end{tabular}
\end{tabular}}

\caption{Demographics: Education level and computer knowledge}
\label{demo_edu_comp2}
\label{demo_education2}
\end{table}

\end{document}